\documentclass[a4paper,12pt]{article}
\usepackage{amsmath,amssymb,amsthm,mathrsfs,graphicx,float,colordvi,url,wrapfig}
\usepackage{here,color,ulem,enumitem,bm,tikz-cd,blkarray,comment}
\usepackage[all]{xy}  
   
\makeatletter

\newcommand{\xleftrightarrow}[2][]{\ext@arrow 3359\leftrightarrowfill@{#1}{#2}} 
\newcommand{\xdashrightarrow}[2][]{\ext@arrow 0359\rightarrowfill@@{#1}{#2}}
\newcommand{\xdashleftarrow}[2][]{\ext@arrow 3095\leftarrowfill@@{#1}{#2}}  
\newcommand{\xdashleftrightarrow}[2][]{\ext@arrow 3359\leftrightarrowfill@@{#1}{#2}} 
\def\rightarrowfill@@{\arrowfill@@\relax\relbar\rightarrow}
\def\leftarrowfill@@{\arrowfill@@\leftarrow\relbar\relax}
\def\leftrightarrowfill@@{\arrowfill@@\leftarrow\relbar\rightarrow}
\def\arrowfill@@#1#2#3#4{
$\m@th\thickmuskip0mu\medmuskip\thickmuskip\thinmuskip\thickmuskip
\relax#4#1 
\xleaders\hbox{$#4#2$}\hfill
#3$
}     
\makeatother
 
\begin{document}
\renewcommand{\thefootnote}{\fnsymbol{footnote}} 
\begin{titlepage}

\vspace*{5mm}

\begin{center}

{\large 
\textbf{Canonical Quantization of the U(1) Gauge Field}
\\[1.0mm]
\textbf{in the right Rindler-wedge in the Rindler Coordinates}\\[1.0mm]
}

\vspace*{7.0mm}

\normalsize
{\large Shingo Takeuchi
}
\vspace*{4.0mm} 

\textit{
\small Faculty of Environmental and Natural Sciences, Duy Tan University, Da Nang, Vietnam}\\
\vspace*{5mm}  
\end{center}

\begin{abstract}
In this study, 
the canonical quantization of the U(1) gauge field 
in the Lorentz-covariant gauge
in the right Rindler-wedge (RRW) of the four-dimensional Rindler coordinates 
is performed.   
Specifically, we obtain the gauge-fixed Lagrangian 
by the Lorentz-covariant gauge in the RRW of the Rindler coordinates, 
which is composed of the U(1) gauge field and  $B$-field. 
Then, we obtain the mode-solutions of the U(1) gauge field and $B$-field 
by solving the equations of motion obtained from that gauge-fixed Lagrangian.
Subsequently, defining the Klein-Gordon inner-product in the RRW of the Rindler coordinates, 
we determine the normalization constants of all directions of the mode-solutions of the U(1) gauge field and $B$-field. 
Then, 
for the U(1) gauge field given by those normalized mode-expanded solutions, 
we obtain the commutation relations of the creation and annihilation operators defined in the RRW of the Rindler coordinates 
by formulating the canonical commutation relations. 
In addition, we provide a polarization vector for the annihilation operators obtained in this way. 

Using these result, we show that the Minkowski ground state can be expressed
as the outer-product of the left and right Rindler-wedges state 
on which those creation and annihilation operators act. 
Then, tracing out the left Rindler states of that Minkowski ground state, 
we obtain the density matrix of the U(1) gauge field in the RRW.
From this, we show that the U(1) gauge field in a constant accelerated system will feel the Unruh temperature as well.
\end{abstract}
\end{titlepage}

\newpage  
  
\allowdisplaybreaks 
\setcounter{footnote}{0}

\section{Introduction}  
\label{sec:int}   


From the analysis of the uniformly accelerated motion,
the Unruh temperature is derived as 
$T_U = \hbar a/(2\pi c k_{\rm B}) \approx 4 \times 10^{-23} \, a/
$(cm/$s^2$)[K]~\cite{Fulling:1972md,Davies:1974th,Unruh:1976db}.
Since the Unruh temperature we can currently produce is less than $3$[K] CMB (Cosmic Microwave Background), 
its detection is beyond our current technological capabilities, 
and the possibility of its detection is a technological issue for us, currently. 
Meanwhile, it is also important as the experimental confirmation of the Hawking radiation. 
Upon reviewing the recent literature on detecting
the Unruh temperature in the experimental field, 
the following studies were retrieved:~Bose-Einstein Condensate~\cite{Retzker}, 
neutrino oscillation~\cite{Luciano:2021onl,Cozzella:2018qew,Dvornikov:2015eqa}, 
anti-Unruh effect~\cite{
Pan:2023tqb,Pan:2021nka,Chen:2021evr,Zhou:2021nyv,Barman:2021oum,Li:2018xil,Garay:2016cpf,Liu:2016ihf,Brenna:2015fga},
cold atoms~\cite{Kosior:2018vgx,Rodriguez-Laguna:2016kri}, 
Berry phases~\cite{Quach:2021vzo,Martin-Martinez:2010gnz}, 
Casimir effect~\cite{Lin:2018wxu,Marino:2014rfa}, 
classical analog~\cite{Leonhardt:2017lwm} 
and others~\cite{Lynch:2019hmk,Lochan:2019osm}. 


Theoretically, the coordinates of a constant accelerated motion are generalized to the Rindler coordinates, in which the Killing horizons exist.
As a result, the thermal excitation analogues to the Hawking radiation exists in the Rindler coordinates, 
and the Rindler coordinates can be regarded as a finite temperature system with the temperature given by the Unruh temperature.  
The prediction of this thermal behavior stimulates the following studies:
analysis of the critical temperatures of the phase transitions~\cite{
Ohsaku:2004rv,Ebert:2006bh,Castorina:2007eb,Castorina:2012yg,Takeuchi:2015nga} 
and the Hagedorn transition in strings~\cite{Parentani:1989gq,Lowe:1994ah}, 
in the Rindler coordinates;
analysis of the thermal radiation from a particle performing a constant accelerated motion~\cite{
Schutzhold:2006gj,Schutzhold:2009scb}\cite{
Iso:2010yq,Iso:2013sm,Oshita:2015xaa,Oshita:2015qka,Iso:2016lua,Yamaguchi:2018cqw}; 
analysis of the Schwinger effect in a constant accelerated system and its application to black hole spacetimes~\cite{
Kim:2016dmm,Kaushal:2022las}; 
and analysis of quantum corrections in the energy-momentum tensors of gases in a constant accelerated system~\cite{
Prokhorov:2019cik,Prokhorov:2019hif,Prokhorov:2019yft}. 
Fundamental issues in the Rindler coordinates have also been investigated:
Unruh radiation in terms of the tunneling~\cite{Kim:2007ep};  
inversion between the bosonic and fermionic statistics occurring in the odd dimensional Rindler coordinates~\cite{
Terashima:1999xp}; 
Unruh temperature in the AdS space~\cite{Deser:1997ri};
and the upper bound for the acceleration~\cite{Rovelli:2013osa}.
Via the Killing horizon included within the Rindler coordinates, 
the Rindler coordinates are exploited to examine 
Einstein's equation as a state equation~\cite{Jacobson:1995ab}, 
the generalized second law~\cite{Wall:2010cj}, and
Rindler-AdS/CFT~\cite{Parikh:2012kg,Fareghbal:2014oba,Sugishita:2022ldv}. 
Also, via the Killing horizon included within the Rindler coordinates, 
the modular Hamiltonians and entanglement are investigated in~\cite{Casini:2008cr,Arias:2016nip,Arias:2017dda}. 
\newline


According to the literature up through the present,
the canonical quantization (CQ) of the gauge fields 
in the Rindler coordinates has not been properly conducted yet,
while that of scalar and spinor field has been conducted in \cite{Higuchi:2017gcd} and \cite{Soffel:1980kx,Ueda:2021nln}, respectively.  
By thoroughly checking references, 
\cite{Moretti:1996zt,Higuchi:1992td,Lenz:2008vw,Zhitnitsky:2010ji,
Soldati:2015xma,Blommaert:2018rsf}
have been found to address the U(1) gauge field in the Rindler coordinates. 
However, in these studies, 
{\bf 1)} it is unclear if the mode-solutions have been obtained by solving the equations of motion, and 
{\bf 2)} the normalization constants (NC) of those contain some speculation and have not been correctly given, 
or their discussion proceeds without the NC.
This point is discussed in Sec.~\ref{sbdeb}, 
comparing the normalized mode-solutions obtained in this study and previous studies.

The reason for the NC having not been correctly determined until now is likely that 
the integrals in Appendix\,\ref{r3g67kb} in this study could not be performed until now.

Since those integrals appear in the Klein-Gordon (KG) inner-products 
between the mode-solutions of the U(1) gauge field in the Rindler coordinates, 
the fact those integrals cannot be performed leads to the following situations; 
{\bf 1)} the NC of the mode-solutions cannot be determined.
{\bf 2)} Accordingly, although the equal-time canonical commutation relations (CCR) of the fields can be formally formed, 
the CCR in terms of the mode-solutions cannot be robustly formulated.
{\bf 3)} Furthermore, analysis performed with the KG inner-products, 
such as taking out of the commutation relations for the creation and annihilation operators, 
cannot be done without ambiguities or speculation.

Considering this situation
(where it is unclear whether the mode-solutions have been obtained or not by solving the equations of motion
and their NC have not been correctly given), 
we will solve the equations of motion for the U(1) gauge field modes in all directions of the RRW of the Rindler coordinates in a very explicit way in this study.
Then, using the integrals needed to perform the KG inner-product between the mode-solutions of the U(1) gauge field as shown in Appendix\,\ref{r3g67kb}, 
we explicitly determine the NC of those mode-solutions in the right Rindler-wedge (RRW) of the Rindler coordinates 
(where we take the Lorentz-covariant gauge). 

At this point, in the process to solve the equations of motion to obtain the mode-solutions, we put an ansatz. 
In this sense, the mode-solutions we obtain are solutions but not general solutions. 
The details concerning this point are described in Sec.\,\ref{f2v4t}.
However, in this study, all directions of the gauge field will be solved by very clearly, 
and no reference has been found in other studies in which all directions of the gauge field are solved in such a clear manner~\cite{Higuchi:1992td,Moretti:1996zt,Lenz:2008vw,Zhitnitsky:2010ji,Soldati:2015xma,Blommaert:2018rsf}  
(more details concerning these points will be discussed in Sec.\,\ref{sbdeb}). 
which would be the advantage of our mode-solutions when compared with those in other studies. 
Therefore, it is believed that there is usefulness in the mode-solutions obtained in this study. 

Next, based on those explicitly obtained normalized mode-solutions, 
we will formulate the CCR of the U(1) gauge field in the RRW of the Rindler coordinates, 
from which we will obtain the commutation relations of the creation and annihilation operators of the U(1) gauge field in the RRW of the Rindler coordinates. 
Then, for the annihilation operators obtained in this way, we will provide a polarization vector. 

Moreover, we will show that the Minkowski ground state can be given
as the outer-product of the left and right Rindler-wedge states 
excited by the creation and annihilation operators of the U(1) gauge field 
in the Rindler coordinates. 
Then, obtaining the density matrix of the U(1) gauge field in the RRW
by tracing out its left Rindler-wedge states,  
we will show that the U(1) gauge field in a constant accelerated system will feel the Unruh temperature as well.
\newline

Regarding this paper's organization, in Sec.\,\ref{vtiss}, 
the Rindler coordinates used in this study are reviewed. 
In Sec.\,\ref{vesbv}, the CQ of the U(1) gauge field 
in the Lorentz-covariant gauge 
in the RRW of the Rindler coordinates is performed.   
In terms of each subsection, in Sec.\,\ref{f367vo}, a gauge-fixed Lagrangian in the Lorentz-covariant gauge is obtained.
In Sec.\,\ref{f2v4t} and \ref{brpnevt}, 
the mode-solutions and the NC of those are obtained. 
Here, the mode-solutions obtained  are not general solutions as mentioned above, 
which is commented on at the end of Sec.\,\ref{f2v4t}.
In Sec.\,\ref{sbdeb}, the normalized mode-solutions obtained in this study are compared with those in  other studies.
In Sec.\,\ref{bywbd}, the CQ is performed, and the commutation relations of the creation and annihilation operators are obtained. 
For the annihilation operators obtained in this way, 
a typical polarization vector for ($S$,$L$,$\pm$)-direction is provided in Sec.\,\ref{yervd}.
As a result, some constraint is derived for the coordinate. 
We discuss its origin at the end of Sec.\,\ref{yervd}.

In Sec.\,\ref{utempu1rc}, 
the density matrix of the U(1) gauge field in the RRW is obtained. 
Regarding each subsection, 
in Sec.\,\ref{usdvea} and \ref{ugfred}, 
${\bm a}^{{\rm (M)}\perp}_{\vec{q}}$ is obtained in terms of 
${\bm a}^{{\rm (R)}\perp}_{\vec{q}}$ and 
$({\bm a}^{{\rm (R)}\perp}_{\vec{q}})^\dagger$ 
by analyzing the Bogoliubov coefficients 
(${\bm a}^{{\rm (M)}\perp}_{\vec{q}}$ 
and ${\bm a}^{{\rm (R)}\perp}_{\vec{q}}$ mean the annihilation operators 
in the Minkowski and Rindler coordinates).
In Sec.\,\ref{eppwte}, from the condition ${\bm a}^{{\rm (M)}\perp}_{\vec{q}}\vert 0_{\rm M} \rangle =0$, 
the ground state in the Minkowski coordinates is constituted by the Rindler states, 
from which in Sec.\,\ref{yedvsh}, the density matrix is obtained, and it is shown that 
the U(1) gauge field in a constant accelerated system will feel the Unruh temperature as well.

In Sec.\,\ref{Summary}, this study is summarized, 
and the future directions for this study are discussed.
In Appendix\,\ref{buobhs}, 
some parts of the analysis in Sec.\,\ref{f367vo} are described as
they are common to the well-known analysis in the Minkowski coordinates.
In Appendix\,\ref{r3g67kb}, 
the integral formulas, 
which play an essential role 
in the calculations of the KG inner-products 
to obtain the NC in Sec.\,\ref{brpnevt},
are given. 

\section{The Rindler coordinates used in this study}
\label{vtiss}

In this section, the Rindler coordinates used in this study are reviewed.

\subsection{$ds^2$ of the LRW and RRW in this study}
\label{vtiss5}

We begin with the $4$-dimensional Minkowski spacetime given as
\begin{eqnarray}\label{vfdbur}
ds^2=c^2dt^2-\sum_{i=1}^3(dx^i)^2.
\end{eqnarray}
We perform the following coordinate transformation:
\begin{subequations}\label{vrbwo0}
\begin{align}
\label{vrbwo}
t =& \,\, c a^{-1}\,e^{ a\xi/c^2} \sinh a\tau, \quad
x^1 = \, c^2 a^{-1}\,e^{ a\xi/c^2 } \cosh a\tau,\\*[1.5mm]
\label{vrbwo2}
t =& \,\, c a^{-1}e^{a\tilde{\xi}/c^2} \sinh a\tilde{\tau}, \quad
x^1 = \, -c^2 a^{-1}e^{a\tilde{\xi}/c^2} \cosh a\tilde{\tau},  
\end{align}
\end{subequations}
where $(\tilde{\tau},\tilde{\xi})$ and $(\tau,\xi)$ are the Rindler coordinates in 
the left and right Rindler-wedges (LRW and RRW), 
in which $a$ is considered to be fixed. 
$\xi$ specifies the trajectory of the constant accelerated motion of the object, 
and $\tau$ parametrizes the trajectory specified by $\xi$, 
which can be identified with the proper time of the constant accelerated motion of the object. 
We take $c$ as $1$ in what follows. 
As a result, (\ref{vfdbur}) can be written as
\begin{eqnarray}\label{astew}
ds^2 =
\left\{ 
\begin{array}{ll}
\! e^{2a\tilde{\xi}} (d\tilde{\tau}^2 -d\tilde{\xi}^2) - (dx^\perp)^2 &
\! \textrm{in LRW}, \\[1.5mm] 
\! e^{2a\xi} (d\tau^2 -d\xi^2) - (dx^\perp)^2 & 
\! \textrm{in RRW},   
\end{array} 
\right.
\end{eqnarray} 
where $x^\perp \equiv (x^2,x^3)$. 
We show the Rindler coordinates $(\tau, \xi)$ and $(\tilde{\tau}, \tilde{\xi})$ 
on the Minkowski coordinates $(t,x)$ in Fig.\ref{wsdd57}.

Let us introduce a coordinate $\rho$ defined as follows: 
\begin{eqnarray}\label{dvyjd}
\rho \equiv 
\left\{
\begin{array}{ll}
\! -a^{-1}e^{a\tilde{\xi}} & \!\textrm{in LRW,} \\[1.5mm] 
\! +a^{-1}e^{a\xi} & \!\textrm{in RRW.} 
\end{array}
\right.
\end{eqnarray}
Then, $ds^2$ for RRW in (\ref{astew}) can be written as
\begin{eqnarray}\label{dres}
ds^2 = a^2\rho^2 d\tau^2-d\rho^2-(dx^\perp)^2.
\end{eqnarray}
$ds^2$ for LRW in (\ref{astew}) is given in the same way but $\tau$ is given by $\tilde{\tau}$. 
Euclideanizing the $\tau$ as $\tau \to -i\tau$ in (\ref{dres}), it can be written as
\begin{eqnarray}\label{hertrt}
ds_{\rm E}^2 = a^2\rho^2 d\tau^2+d\rho^2+(dx^\perp)^2,
\end{eqnarray}
where we denoted $-ds^2$ as $ds_{\rm E}^2$. 
In (\ref{hertrt}), the $\tau$-direction is periodic by $\beta = 2\pi/a$, 
which agrees with the inverse of the Unruh temperature in the uniformly accelerated system with $a$.

\subsection{The light-cone coordinates in the Minkowski coordinates}
\label{vtiss4}

\begin{figure}[H]   
\vspace{0mm} 
\begin{center}
\includegraphics[clip,width=4.5cm,angle=0]{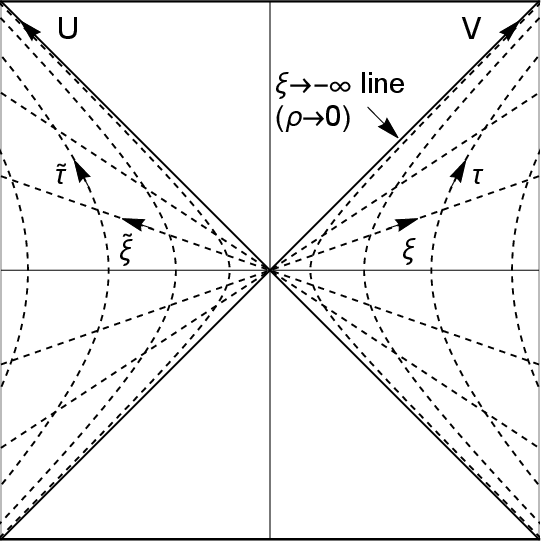} 
\end{center}
\vspace{-5.0mm}
\caption{
This figure represents the Rindler coordinates; 
$(\xi,\tau)$ are defined in (\ref{vrbwo0}),  
$(U,V)$ are the light-cone coordinates defined in (\ref{rebds}), 
and ``the $\xi \to -\infty$ line''  is defined in Sec.\,\ref{iytjeb}, 
which is generally referred to as the Killing horizon.
} 
\label{wsdd57}
\end{figure}

We define the light-cone coordinates 
in the Minkowski coordinates as follows:
\begin{eqnarray}\label{rebds}
U \equiv t-x^1, \quad V \equiv t+x^1. 
\end{eqnarray}
From Fig.{\ref{wsdd57}}, we can see that:
\begin{eqnarray}\label{rvdsd}
U
\left\{
\begin{array}{ll}
\! > 0 & \!\textrm{in LRW,} \\*[1.5mm] 
\! < 0 & \!\textrm{in RRW,} 
\end{array}
\right. 
\quad
V
\left\{
\begin{array}{ll}
\! < 0 & \!\textrm{in LRW,} \\*[1.5mm] 
\! > 0 & \!\textrm{in RRW.} 
\end{array}
\right. 
\end{eqnarray} 
Based on (\ref{vrbwo}) and (\ref{vrbwo2}), we can represent these 
in terms of the LRW and RRW coordinates ($\tilde{\tau}$, $\tilde{\xi}$) and ($\tau$, $\xi$) as follows:
\begin{eqnarray}\label{rerwod}
(U,V) =
\left\{
\begin{array}{ll}
\! a^{-1}( e^{a\tilde{u}},\, -e^{-a\tilde{v}}) & \!\textrm{in LRW,} \\*[1.5mm] 
\! a^{-1} (-e^{-au},\, e^{av}) & \!\textrm{in RRW,} 
\end{array}
\right. 
\end{eqnarray}
where 
$(\tilde{u},\tilde{v}) \equiv (\tilde{\tau}-\tilde{\xi}, \tilde{\tau}+\tilde{\xi})$ and
$(u,v) \equiv (\tau-\xi, \tau+\xi)$. 
We show the coordinates $(U,V)$  in Fig.\ref{wsdd57}.

\subsection{Definition of the Killing horizon}
\label{iytjeb}

Let us consider taking $\xi$ in (\ref{vrbwo}) closer to $-\infty$, 
and refer to the line asymptoted to at that time as ``the $\xi \to -\infty$ line''.  
Note that the limit taking $\xi$ exactly to $-\infty$ is excluded by the definition of the Rindler coordinates.

Then, at any points in $t>0$ or $t<0$ on the $\xi \to -\infty$ line, 
$\tau$ and $\xi$  in (\ref{vrbwo}) should be in the following relation: 
\begin{eqnarray}\label{bekwfq}
\vert\tau\vert \sim \vert \xi \vert. 
\end{eqnarray}
Otherwise, on the $\xi \to -\infty$ line, neither $t$ nor $x^1$ can be finite 
(therefore, the motion of the object in the finite $\tau$ on the $\xi \to -\infty$ line is all packed 
into the neighborhood of $t=0$). 
This can be seen from the forms of $t$ and $x^1$ in (\ref{vrbwo})\footnote{
$t$ and $x^1$ in (\ref{vrbwo}) can be roughly written as
\begin{subequations}\label{wegfy}
\begin{align}
t   &\sim e^{a\xi} \sinh a\tau \sim e^{a(\xi+\tau)}-e^{-a(-\xi+\tau)},\\[1.5mm]
x^1 &\sim e^{a\xi} \cosh a\tau \sim e^{a(\xi+\tau)}+e^{-a(-\xi+\tau)}.
\end{align}
\end{subequations}
Then, on the $\xi \to -\infty$ line, 
if (\ref{bekwfq}) is satisfied, 
$t$ can take finite values as can be seen below:
\begin{eqnarray}
\displaystyle
t \sim 
\left\{
\begin{array}{ll}
e^{a(\xi+\tau)}-e^{-2a\infty}   & \textrm{for $\tau \sim -\xi$,}\\[1.5mm] 
e^{-2a\infty}-e^{-a(-\xi+\tau)} & \textrm{for $\tau \sim +\xi$,} 
\end{array}
\right.
\end{eqnarray}
where $a$ is supposed as some finite value, 
and $\sim$ in ``$\xi \sim \mp\tau$'' mean that 
the values of both sides are same order each other.
By the same logic, $x$ can also take finite values on the $\xi \to -\infty$ line, 
if (\ref{bekwfq}) is satisfied.
}. (\ref{bekwfq}) is used in the analysis in Sec.\,\ref{ugfred}.

Therefore, on the $\xi \to -\infty$ line, 
since $\vert\tau\vert$ asymptotes to $\infty$, 
$t/x^1=\tanh(a\tau)$ closes to $\pm 1$.
Such lines are the $\pm 45$-degree straight diagonal lines in the RRW in Fig.\ref{wsdd57}. 
Here, as the $ds^2$ in the RRW in (\ref{astew}) is invariant for the variation of $\tau$, 
the $\tau$-direction is a Killing vector in the RRW, which we denote as $\partial_\tau$\footnote{
$\partial_\tau$ is given as $a(x^1\partial_t+t\partial_{x^1})$ in the $(t,x^1)$ coordinates in (\ref{vrbwo}).}.
Then, it can be seen that $\partial_\tau$ is the normal vector for those $\pm 45$-degree straight diagonal lines at the same time being the tangent vector; 
therefore, according to the general definition of the null hyeprsurface or Killing horizon,
those $\pm 45$-degree straight diagonal lines are the null hyeprsurface or Killing horizon for $\partial_\tau$.
We employ the Killing horizon as the name.
Then, those $+45$- and $-45$-degree straight diagonal lines are referred to as the future and  past Killing horizon, respectively.

\section{The canonical quantization of the U(1) gauge field in the Rindler coordinates}
\label{vesbv}

In this section, the canonical quantization of the U(1) gauge field in the RRW in the Rindler coordinates is performed in the Lorentz-covariant gauge 
by first  taking the Coulomb gauge as in (\ref{bef67}) and finally taking the Lorentz-covariant gauge as in (\ref{sbrtv}). 
Here, we first  obtain the form of the path-integral in the Coulomb gauge as in (\ref{tedrha}), 
then replace that Coulomb gauge with the Lorentz-covariant gauge. 
Since the process to obtain the form of the path-integral in the Coulomb gauge explicitly depends on the Rindler coordinates,
we perform it in the body text.
On the other hand, 
the replacement of the Coulomb gauge with the Lorentz-covariant gauge can be performed 
in the same way as the case of the Minkowski coordinates and in the coordinate-independent manner;
therefore, we note the analysis to replace the Coulomb gauge with the Lorentz-covariant gauge performed after (\ref{tedrha}) 
in Appendix\,\ref{buobhs}.
The canonical quantization of the U(1) gauge field in the LRW can be immediately obtained 
if that in the RRW is known.

\subsection{The Lagrangian of the U(1) gauge field in the Lorentz-covariant gauge in the Rindler coordinates}
\label{f367vo}

We consider the following Lagrangian density (referred to as Lagrangian) of the U(1) gauge field on the RRW (\ref{dres}): 
\begin{eqnarray}\label{uieqp}
S=\int_{\rm RRW} \! d^4x \,\sqrt{-g}\,{\cal L}_{\rm U(1)}, \quad 
{\cal L}_{\rm U(1)} = -F_{\mu\nu} F^{\mu\nu}/4, 
\end{eqnarray}
where 
$F_{\mu\nu}=\nabla_\mu A_\nu-\nabla_\nu A_\mu=\partial_\mu A_\nu-\partial_\nu A_\mu$ 
($\nabla_\mu$ is covariant derivative given by the metrices in (\ref{dres})). 
The Christoffel symbols in the RRW (\ref{dres}) are given as follows: 
\begin{eqnarray}\label{rmwg}
\Gamma^{0}_{01} = \rho^{-1},\quad
\Gamma^{1}_{00} = a^2\rho, \quad \textrm{others $=0$}.
\end{eqnarray}

It is known that the Lagrangian in the Lorentz-covariant gauge in the Minkowski coordinates is given as follows:
\begin{eqnarray}\label{ebtwf}
T^{(M)}
\!\!\! &\equiv& \!\!\!
\int \! {\cal D}\!A^k \,{\cal D}B  \,{\cal D}c\,{\cal D}\bar{c}\,
\exp[\, i \! \int \! d^4x \, \sqrt{-g} \, 
{\cal L}^{(M)}],
\\*[1.5mm]
{\cal L}^{(M)}
\!\!\! &=& \!\!\!
{\cal L}_{\rm U(1)}
+B \, \partial_\mu A^\mu + B^2/2
+i\,\bar{c} \,\partial_\mu \partial^\mu \, c,  \nonumber
\end{eqnarray}
where $\nabla_\mu A^\mu ={\cal C}$ is taken 
as the Lorentz-covariant gauge (${\cal C}$ is some real function).
In this, replacing the differentials with the covariant derivatives (and $d^4x$ with $d^4x\, \sqrt{-g}$), 
we can obtain the Lagrangian in the Lorentz-covariant gauge in the Rindler coordinates:
\begin{eqnarray}\label{etsiph}
{\cal L}^{(R)}
\!\!\! &=& \!\!\!
{\cal L}_{\rm U(1)}
+B \, \nabla_\mu A^\mu + B^2/2
+i\,\bar{c} \,\nabla_\mu \nabla^\mu \, c.
\end{eqnarray}
Based on this Lagrangian, all fields have the canonical conjugate momentum as follows:
\begin{eqnarray}\label{rr54wh}
&& \bullet \quad
\frac{\partial{\cal L}^{(R)}}{\partial (\nabla_0 A^k)} 
=F_k{}^0 \equiv \pi_k, \quad 
\frac{\partial{\cal L}^{(R)}}{\partial (\nabla_0 A^0)}    = B = -\nabla_\mu A^\mu  \equiv \pi_{0}, \quad 
\nonumber\\*[1.5mm]
&& \bullet \quad \,\,
\frac{\partial{\cal L}^{(R)}}{\partial (\nabla_0 B)}  
=-A^0 \equiv \pi^{(B)}, \quad 
\frac{\partial{\cal L}^{(R)}}{\partial (\nabla_0 c)}       =  i\,\nabla^0\bar{c}   \equiv \pi^{(c)}, \quad 
\frac{\partial{\cal L}^{(R)}}{\partial (\nabla_0 \bar{c})} = -i\,\nabla^0 c        \equiv \pi^{(\bar{c})},
\nonumber\\
\end{eqnarray}
where $k=1,\perp$, and an equation of motion (\ref{e3ier2}) is used in the relation between $B$ and $-\nabla_\mu A^\mu$.
However, in arriving at (\ref{ebtwf}), 
the Coulomb gauge (non-covariant gauge) is used as can be seen in (\ref{vrue}).
Therefore, we will check if (\ref{etsiph}) can be obtained from (\ref{ebtwf}) in the following subsection. 
\newline

Only $A^\mu$ and $B$ are addressed in this study.
Lastly in this subsection, let us comment on the validity of this.  
In the system defined by (\ref{etsiph}), 
the equations of motion of the ghost fields 
and the equations of motion of  $A^\mu$ and $B$ decouple each other, 
which can be read from (\ref{ebtwf}).
As a result, the partition functions of the ghost and the gauge fields can decouple each other, 
such as $Z=Z_{A^\mu,B} \, Z_{c,\bar{c}}$.

The reason for this is that the Faddeev-Popov determinant (FPd) does not include the gauge field,  
if the addressed gauge field is the U(1), as can be seen in (\ref{sbrtv}).
Therefore, since the FPd behaves as a constant in the path-integral with regard to the U(1) gauge field, 
we can suppose that the ghost field had not existed from the beginning.  
(However, the BRST transformation is given by the gauge parameter given by the product of some Grassmann number and the ghost field.)
  
\subsubsection{The Hamiltonian density}
\label{ts6dt5}

The conjugate momenta of $A^k$ and $A_k$  are given from ${\cal L}_{\rm U(1)}$ in (\ref{uieqp}) as follows:
\begin{subequations}\label{nireb}
\begin{align}
\label{nireb1}
\frac{\partial {\cal L}_{\rm U(1)}}{\partial (\partial_0 A^k)}
=& \,\,
F_k{}^0 
= \pi_k
\,\equiv\, E_k, \\*[1.5mm]
\label{nireb2}
\frac{\partial {\cal L}_{\rm U(1)}}{\partial (\partial_0 A_k)}
=& \,\,
F^{k 0} 
= \pi^k
\,\equiv\, E^k,
\end{align}
\end{subequations}
where $\pi_k$ are defined in (\ref{rr54wh}), 
and $E_k$ and $E^k$ mean the electric fields; 
$k=1, \perp$ in the RRW (\ref{dres}).

Using $E_k$ and $E^k$ defined in (\ref{nireb}), 
${\cal L}_{\rm U(1)}$ in (\ref{uieqp}) can be given as follows:
\begin{eqnarray}\label{bv64r}
{\cal L}_{\rm U(1)}
\!\!\! &=& \!\!\! 
F_{\mu 0}F^{\mu 0}/2-F_{ij}F^{ij}/4
\nonumber\\*[1.5mm]
\!\!\! &=& \!\!\! 
-E^k (\nabla_k A_0-\nabla_0 A_k)-(B_k^2-g_{00}E_k^2)/2,
\end{eqnarray}
where $B^k \equiv -\varepsilon^{ijk}F_{ij}/2$, $B_k \equiv \varepsilon_{ijk}F^{ij}/2$ and
$\varepsilon^{ijk}=-\varepsilon_{ijk}$  in the RRW (\ref{dres}) ($B_k$ and $B^k$ mean the magnetic fields). 
$B_k^2=B_k B^k$ ($E_k^2$ is likewise).

The first term in the r.h.s. of (\ref{bv64r}) can be rewritten as
\begin{eqnarray}\label{erhd}
\int_{\rm RRW} \! d^4x \sqrt{-g}\,E^{k} \,\nabla_k A_0
= -\int_{\rm RRW} \! d^4x \sqrt{-g}\,\partial_k E^k A_0, 
\end{eqnarray}
where in the rewritten above, 
expressing $E^k \, \nabla_k A_0$ as $F^k{}_0 \,\nabla_k A^0$ ($A_0$ has been changed to $A^0$), 
$\nabla_k A^0$ has been rewritten as $(\sqrt{-g})^{-1} \partial_k (\sqrt{-g} A^0)$ 
assuming the boundary condition that 
the fields vanish at the infinite far region which is given by $\xi$ to $\infty$.  
This rewriting is performed in order to change the coefficients of $A_0$, which is the operator $\nabla_k$, 
to numbers for the convenience in the path-integral with regard to $A^0$ in (\ref{w5jsk5}).

With the rewriting (\ref{erhd}), 
the Lagrangian  (\ref{bv64r}) can be given as
\begin{eqnarray}\label{b65urfb}
\textrm{(\ref{bv64r})}
=
\partial_k E^k A_0+ E^{k} \,\nabla_0 A_k
-(B_k^2-g_{00}E_k^2)/2. 
\end{eqnarray}
With the Lagrangian given by (\ref{b65urfb}), 
the Hamiltonian density (referred to as the Hamiltonian) can be obtained as follows:
\begin{eqnarray}\label{bvery}
{\cal H}
=\nabla_0 A_\mu\,\pi^\mu
-{\cal L}_{\rm U(1)}
=
\nabla_0 A^0 \pi_0
-\partial_k E^k A_0
+(B_k^2-g_{00}E_k^2)/2. 
\end{eqnarray}

\subsubsection{Constitution of the path-integral}
\label{tduk24}

In the system with the Hamiltonian (\ref{bvery}), 
there are two constraint conditions $\phi^{(i)}$ $(i=1,2)$:
\begin{eqnarray}\label{bef67} 
\phi^{(1)} \equiv \pi_0 = 0,\quad
\phi^{(2)} \equiv \partial_k E^k = 0.
\end{eqnarray}
These $\phi^{(i)}$ $(i=1,2)$ form the first-class constraint. 
Corresponding to these two conditions of the first-class constraint, 
we take the Coulomb gauge $\chi^{(i)}$ as follows:
\begin{eqnarray}\label{vrue}
\chi^{(1)}\equiv A^0=0, \quad 
\chi^{(2)}\equiv \nabla_k A^k=0.
\end{eqnarray}
Denoting these together as 
$\tilde{\phi} \equiv \{\phi^{(1)},\phi^{(2)},\chi^{(1)},\chi^{(2)}\}$, 
the equal-time Poisson bracket (referred to as the Poisson bracket) for $\tilde{\phi}$ can be obtained for each $\tau$ 
as follows:
\begin{equation}\label{ivw3rk}
\big[\{\tilde{\phi}(x),\tilde{\phi}(y)\}_{\rm P.B.}\big]=
\left[
\begin{array}{cccc}
0 & 0 & -1 & 0                       \\
0 & 0 & 0  & -\nabla^k\nabla_k       \\
1 & 0 & 0  & 0                       \\
0 & \nabla^k\nabla_k & 0  & 0 
\end{array}
\right]
\times \delta^3(\vec{x}-\vec{y}). 
\end{equation}
Here, the Poisson bracket are defined as follows:
\begin{subequations}
\begin{align}
\{X(\tau,\vec{x}),Y(\tau,\vec{y})\}_{{\rm P.B.}}
&= 
\left(\frac{\partial X(\tau,\vec{x})}{\partial A^\mu(\tau,\vec{x})} \frac{\partial Y(\tau,\vec{y})}{\partial \pi_\mu(\tau,\vec{y})}
- \frac{\partial X(\tau,\vec{x})}{\partial \pi_\mu(\tau,\vec{y})} \frac{\partial Y(\tau,\vec{x})}{\partial A^\mu(\tau,\vec{y})}\right)
\delta^3(\vec{x}-\vec{y}), 
\end{align}
\end{subequations}
where $X(\tau,\vec{x})$ and $Y(\tau,\vec{y})$ are some functions, 
and $\delta^3(\vec{x}-\vec{y})=\displaystyle \delta(x^1-y^1)\,\delta^2(x^\perp-y^\perp)$ in the RRW (\ref{dres});
$\pi_\mu$ have been defined in (\ref{rr54wh}).
Since ${\rm det} \{\tilde{\phi}(x),\tilde{\phi}(y)\}_{\rm P.B.}$ in (\ref{ivw3rk}) is non-vanishing, 
$\tilde{\phi}$ forms the second-class constraint. 
When the conditions (\ref{bef67}) and (\ref{vrue}) are imposed 
in the phase space of the $(A^\mu,\pi_\mu)$, 
the path-integral for the U(1) gauge field in the RRW (\ref{dres}) can be written as follows:
\begin{eqnarray}\label{iedub}
T^{(R)}
\!\!\! &\equiv& \!\!\!
\int \! {\cal D}\!A \, {\cal D}\pi 
\prod_{x \in {\rm RRW}}
\big[\delta(\phi^{(1)})\,\delta(\phi^{(2)})\,\delta(\chi^{(1)})\,\delta(\chi^{(2)})\big]\cdot
\prod_{\tau} 
\textrm{Det}\big[ M_c(x,y)\big]
\nonumber\\*
&& \times 
\exp \big[ 
i \int_{\rm RRW} \! d^4x \,\sqrt{-g}\,\{ \nabla_0 A^\mu\,\pi_\mu-{\cal H}\}
\big],
\end{eqnarray}
where 
\begin{subequations}\label{neic}
\begin{align}
\label{neic1}
M_c(x,y) 
&\equiv 
\big[\{\phi(x),\chi(y)\}_{\rm P.B.}\big]
=\nabla^k \nabla_k \delta^3(\vec{x}-\vec{y}), 
\\*[1.5mm]
\label{neic2}
{\cal D}\!A \, {\cal D}\pi 
&\equiv  
\prod_{\mu=0}^3\prod_{x \in {\rm RRW}} dA^\mu(x)\,d\pi_\mu(x), 
\end{align}
\end{subequations}
where
$\phi \equiv \{ \phi^{(1)},\phi^{(2)} \}$ and $\chi \equiv \{ \chi^{(1)},\chi^{(2)} \}$. 

Here, there is an issue of $A^\mu$ and $A_\mu$ and $\pi^\mu$ and $\pi_\mu$, 
which we take as the variable to be path-integrated, respectively. 
As for this issue, 
we take $A^\mu$ and $\pi_\mu$, as can be seen in (\ref{neic2}). 
Therefore, $A_\mu$ and $\pi^\mu$ are considered as $g_{\mu\nu}A^\nu$ and $g^{\mu\nu}\pi_\nu$ 
in the path-integral in the following subsection 
($g_{\mu\nu}$ are the metrices of the RRW (\ref{dres})).

\subsubsection{Constitution of the Lagrangian in the Lorentz-covariant gauge}
\label{teeah54}

Let us integrate out $\pi_0$ in (\ref{iedub}), 
which can be performed readily as $\phi^{(1)}=\pi_0$ as in (\ref{bef67}) 
and ${\cal L}_{\rm U(1)}$ does not include $\pi_0$, where $\nabla_0 A^k\,\pi_k-{\cal H}={\cal L}_{\rm U(1)}$.
Therefore, now (\ref{iedub}) can be written as follows:
\begin{eqnarray}\label{ssrrt}
\textrm{(\ref{iedub})}
\!\!\! &=& \!\!\!
\int \! {\cal D}\!A \, {\cal D}\pi_{k}\, 
\prod_{x \in {\rm RRW}}
\big[\delta(\chi^{(1)})\,\delta(\chi^{(2)})\,\delta(\phi^{(2)})\big]\cdot 
\prod_{\tau} \textrm{Det}\big[M_c(x,y)\big]\nonumber\\*
&& \!\!\!\! \times\,  
\exp \big[
i \int_{\rm RRW} d^4x \,\sqrt{-g}\,  
{\cal L}_{\rm U(1)}|_{\pi_0=0}
\big].
\end{eqnarray}
Since ${\cal L}_{\rm U(1)}$ does not include $\pi_0$, 
we can write ${\cal L}_{\rm U(1)}|_{\pi_0=0}$ just as ${\cal L}_{\rm U(1)}$ in the following.
 
Then, introducing the new functional variable $\eta=\eta(x)$, 
we can give $\prod_{x \in {\rm RRW}} \delta(\phi^{(2)})$ by the functional integral:
\begin{eqnarray}\label{bvfer} 
\prod_{x \in {\rm RRW}} \delta(\phi^{(2)})
=\int {\cal D}\eta \, \exp \big[ i \int_{\rm RRW} d^4x \,\sqrt{-g}\,\eta \, \phi^{(2)} \big]. 
\end{eqnarray} 
Therefore, (\ref{ssrrt}) can be written as follows:
\begin{eqnarray}\label{vndrtf}
\textrm{(\ref{ssrrt})}
\!\!\! &=& \!\!\!
\int \! {\cal D}\!A \, {\cal D}\pi_{k}
\int {\cal D} \eta \, 
\prod_{x \in {\rm RRW}}
\big[\delta(\chi^{(1)})\,\delta(\chi^{(2)})\big]\cdot 
\prod_{\tau} \textrm{Det}\big[M_c(x,y)\big]\nonumber\\*
&& \!\!\!\! \times\,  
\exp \big[
i \int_{\rm RRW} d^4x \,\sqrt{-g}\, \{ 
{\cal L}_{\rm U(1)}
+\eta \, \partial_k E^k \}
\big],
\end{eqnarray}
where $\nabla_0 A^k\,\pi_k-{\cal H}={\cal L}_{\rm U(1)}$ and $\phi^{(2)}=\partial_k E^k$.

Now let us integrate out $A^0$. 
By this, since $\chi^{(1)}=A^0$ as in (\ref{vrue}), 
the first term in the r.h.s. in ${\cal L}_{\rm U(1)}$ given by (\ref{b65urfb}), 
which is $\partial_k  E^kA_0=\partial_k  E^k g_{00}A^0$, disappears. 
However, since $\eta$ is the variable in the path-integral, 
we can take $\eta$ as $A_0$,
by which 
the term $\eta \, \partial_k  E^k$ 
in (\ref{vndrtf}) 
becomes the  disappeared term $\partial_k  E^k A_0$, 
and ${\cal L}_{\rm U(1)}$ in (\ref{vndrtf}) can revive as it was; namely: 
\begin{eqnarray}\label{w5jsk5}
{\cal L}_{\rm U(1)}+\eta \, \nabla^k \pi_k
\!\!\! &\xrightarrow[\,\, A^0 \,\to\, 0 \,\,]{}& \!\!\!
[{\cal L}_{\rm U(1)}+\eta \, \nabla^k \pi_k] 
\,\big\vert_{A^0 \,\to\, 0}
\nonumber\\*[1.5mm]
\!\!\! &\xrightarrow[\,\,\eta \,\to\, A_0\,\,]{}& \!\!\!
[{\cal L}_{\rm U(1)}+\eta \, \nabla^k \pi_k] 
\,\big\vert_{A^0 \,\to\, 0, \, \eta \,\to\, A_0}
= \textrm{${\cal L}_{\rm U(1)}$ in (\ref{b65urfb})},
\end{eqnarray}
where since $g_{00}$ is constant, 
in ${\cal D} \eta\vert_{\eta \to A_0} = {\cal D} (g_{00}A^0)$,
the contribution of $g_{00}$ has been gotten out as some constant and ${\cal D} (g_{00}A^0)$ has been regarded as ${\cal D} A^0$.
As a result, we can  write (\ref{vndrtf}) as follows:
\begin{eqnarray}\label{vtiof}
\textrm{(\ref{vndrtf})}
\!\!\! &=& \!\!\!
\int \! {\cal D}\!A \,  {\cal D}\pi_{k}
\prod_{x \in {\rm RRW}}
\big[\delta(\chi^{(2)})\big] \cdot
\prod_{\tau} \textrm{Det}\big[M_c(x,y)\big]
\nonumber\\*
&& \!\!\!\! \times\,  
\exp \big[
i \int_{\rm RRW} d^4x \,\sqrt{-g}\, {\cal L}_{\rm U(1)} 
\big].
\end{eqnarray}
 
Now, ${\cal L}_{\rm U(1)}$ in (\ref{vtiof}) is given by (\ref{b65urfb}). 
Performing the rewriting from (\ref{bv64r}) to (\ref{b65urfb}) with (\ref{erhd}) in reverse, 
${\cal L}_{\rm U(1)}$ in (\ref{vtiof}) can be given by the quadratic form with regard to $\pi_k$ as
\begin{eqnarray}\label{h4h6}
\textrm{${\cal L}_{\rm U(1)}$ in (\ref{vtiof})}
\!\! &=& \!\!
F_{0k}\,\pi^k 
-(B_k^2-g_{00}\,\pi_k^2)/2 
\nonumber\\*[1.5mm]
\!\! &=& \!\!
-\frac{g_{00}}{2}
\sum_{k}
(\pi_k+g^{00}F_{0k})
(\pi_k+g^{00}F_{0k})
+\frac{1}{2}(g^{00}\sum_{k}F_{0k}F_{0k}-B_k^2),
\end{eqnarray}
where, since we have set index $k$ to the subscript (namely, not used the Einstein convention), 
we have written the symbol of the summation.
We can integrate out $\pi_k$ as the Gaussian integral, 
and $(g^{00}\sum_{k}F_{0k}F_{0k}-B_k^2)/2={\cal L}_{\rm U(1)}$, 
where if $A^\mu$ and $\pi_\mu$ were not independent of each other, and were in the relation (\ref{nireb1}), 
the r.h.s. of (\ref{h4h6}) becomes $0+{\cal L}_{\rm U(1)}$.
As a result, we can write (\ref{vtiof}) as
\begin{eqnarray}\label{tedrha}
\textrm{(\ref{vtiof})}
=
\int \! {\cal D}\! A
\prod_{x \in {\rm RRW}}
\big[\delta(\chi^{(2)})\big] \cdot
\prod_{\tau} \textrm{Det}\big[M_c(x,y)\big]
\cdot
\exp \big[
i \int_{\rm RRW} d^4x \,\sqrt{-g}\, {\cal L}_{\rm U(1)}
\big].
\end{eqnarray}
From this, we can obtain (\ref{ebtwf})
by proceeding in the same way as the well-known case of the Minkowski coordinates, 
which can be performed in the coordinate-independent manner 
unlike the analysis until which (\ref{tedrha}) has been obtained.
Therefore, we perform the analysis from (\ref{tedrha}) to (\ref{etsiph}) in Appendix\,\ref{buobhs}. 

\subsection{The mode-solutions of the U(1) gauge field in the RRW}
\label{f2v4t} 

In this subsection, 
solving the equations of motion obtained from the gauge-fixed Lagrangian in (\ref{ebtwf}),
the classical mode-solutions of the U(1) gauge field in the RRW (\ref{dres}) are obtained. 
The results are noted in (\ref{ba54nwy}),  
and the fields as the solution are noted in (\ref{w4aeea}). 
How we have solved the equations of motion is noted under (\ref{w4aeea}). 
The mode-solutions obtained in this study are not the general solution, 
which we comment on at the end of this subsection.
\newline

From (\ref{etsiph}), the equations of motion can be obtained as follows\footnote{
$\int d^4x \sqrt{-g}\,{\cal L}^{(R)} = \int d^4x \sqrt{-g}\,(\cdots -A^\mu\, \partial_\mu B + \cdots)$ 
and 
$\partial_\mu (\sqrt{-g}\,F^{\mu \nu})=\sqrt{-g} \,\nabla_\mu F^{\mu \nu}$. 
}:
\begin{subequations}\label{e3ier}
\begin{align}
\label{e3ier1}
-\nabla_\mu F^{\mu \nu}+\partial^\nu B \, &= \, 0, \\*[1.5mm]
\label{e3ier2}
\nabla_\mu A^\mu + B \, &= \, 0.
\end{align}
\end{subequations} 
From these, equations of motion for the fields in the RRW (\ref{dres}) are obtained as follows: 
\begin{subequations}\label{ebraue}
\begin{align}
\label{ebraue1}
(g^{\mu\nu}\partial_\mu \partial_\nu-3\rho^{-1}\partial_1)A^0 
&= -2a^{-2}\rho^{-3}\partial_0 A^1, 
\\*[1.5mm]
\label{ebraue2}
(g^{\mu\nu}\partial_\mu \partial_\nu-3\rho^{-1}\partial_1-\rho^{-2})A^1 
&= 2\rho^{-1}(B+\partial_\perp A^\perp), \\*[1.5mm]
\label{ebraue3}
(g^{\mu\nu}\partial_\mu \partial_\nu-\rho^{-1}\partial_1)A^\perp &=0, 
\\*[1.5mm]
\label{ebraue4}
\partial_0 A^0 +\partial_1 A^1+\partial_\perp A^\perp+\rho^{-1}A^1 +B &=0, 
\\*[1.5mm]
\label{ebraue5}
(g^{\mu\nu}\partial_\mu \partial_\nu-\rho^{-1}\partial_1)B &=0,
\end{align}
\end{subequations}
where 
$0$- and $1$-direction mean $\tau$- and $\rho$-direction respectively, 
and $\perp$ means $2$- and $3$-directions, in (\ref{dres}).
Non-zero $\Gamma^\mu_{\nu \lambda}$ have been noted in (\ref{rmwg}). 
Combining (\ref{e3ier1}) and (\ref{e3ier2}), 
$\nabla_\nu\nabla^\nu A^\mu=0$ can be obtained, 
from which (\ref{ebraue1})-(\ref{ebraue3}) can be obtained 
(in obtaining (\ref{ebraue2}), (\ref{ebraue4}) is used). 
From (\ref{e3ier2}), (\ref{ebraue4}) can be obtained.
Multiplying the entire (\ref{e3ier1}) by $\nabla_\nu$, 
$\nabla_\mu \nabla^\mu B=0$ can be obtained, from which (\ref{ebraue5}) can be obtained.

Since the coordinate system in the RRW (\ref{dres}) is homogeneous for the $(0,\perp)$-direction as can seen in (\ref{dres}), 
the fields in the RRW can be given by the following Fourier expansion:
\begin{subequations}\label{wyea}
\begin{align}
\label{wyea1}
A^\mu(\tau,\rho,x^\perp) &=
\int_{-\infty}^\infty \! dk_0 \int_{-\infty}^\infty \! d^{2}k_\perp\,
\tilde{{\cal N}}^{(\mu)}_{k}
\tilde{A}^\mu_{k}(\rho) \, 
e^{-ikx}, \\*[1.5mm]
\label{wyea2}
B(\tau,\rho,x^\perp) &=
\int_{-\infty}^\infty \! dk_0 \int_{-\infty}^\infty \! d^{2}k_\perp\,
\tilde{{\cal N}}^{(B)}_{k}
\tilde{B}_{k}(\rho) \,
e^{-ikx}, 
\end{align}
\end{subequations}
where
\begin{itemize}
\item[$\cdot$]
$k$ and $kx$ in the subscripts and shoulder of $e$ in r.h.s. 
are abbreviations of ``$k_0,k_\perp$'' and ``$k_0\tau-k_\perp x^\perp$'', respectively 
(in this study, we use these notations).   

\item[$\cdot$]
$A^\mu(\tau,\rho,x^\perp)$ and $B(\tau,\rho,x^\perp)$ in this study are assumed to be real.

\item[$\cdot$]
$\tilde{{\cal N}}^{(\mu)}_{k}$ and $\tilde{{\cal N}}^{(B)}_{k}$ are constant of each mode, 
which can take complex numbers. 
We decompose these into the normalization constant part 
and some coefficient part irrelevant of the normalization constant part. 
\begin{eqnarray}\label{rwfyg53}
\tilde{{\cal N}}^{(\mu)}_{k}=\, {\cal N}^{(\mu)}_{k}\,\bm{a}^{(\mu)}_{k}, \quad
\tilde{{\cal N}}^{(B)}_{k}  =\, {\cal N}^{(B)}_{k}\,\bm{b}_{k}, 
\end{eqnarray}
where ${\cal N}^{(\mu)}_{k}$ and $ {\cal N}^{(B)}_{k}$ mean the normalization constant part 
and $\bm{a}^{(\mu)}_{k}$ and $\bm{b}_{k}$ mean some coefficient part, 
which will become the annihilation operator 
when the canonical quantization is performed as seen in Sec.\,\ref{bywbd}.
\end{itemize}

Applying (\ref{wyea}) to (\ref{ebraue}), 
the following equations of motion in terms of the modes can be given as follows:
\begin{subequations}\label{r4rg2}
\begin{align}
\label{r4rg21}
(
(a\rho)^{-2}k_0^2
+\partial_1^2
-k_\perp^2
+3\rho^{-1}\partial_1)     \,\tilde{\cal N}^{(0)}_{k}\tilde{A}^0_{k}
&=- 2a^{-2}\rho^{-3}\,ik_0 \,\tilde{\cal N}^{(1)}_{k}\tilde{A}^1_{k}, 
\\*[1.5mm]
\label{r4rg22}
(
\rho^{-2}(a^{-2}k_0^2+1)
+\partial_1^2
-k_\perp^2
+3\rho^{-1}\partial_1) 
\,\tilde{\cal N}^{(1)}_{k}\tilde{A}^1_{k}
&= -2\rho^{-1}(\tilde{{\cal N}}^{(B)}_{k} \tilde{B}_{k}+ik_\perp \,
\tilde{\cal N}^{(\perp)}_{k} \tilde{A}^\perp_{k}),  
\\*[1.5mm]
\label{r4rg23}
((a\rho)^{-2}k_0^2+\partial_1^2-k_\perp^2+\rho^{-1}\,\partial_1)
\,\tilde{\cal N}^{(\perp)}_{k}\tilde{A}^\perp_{k}
&=0, 
\\*[1.5mm]
\label{r4rg25}
-ik_0 \,\tilde{\cal N}^{(0)}_{k}\tilde{A}^0_{k} 
+ (\partial_1+\rho^{-1})
\,\tilde{\cal N}^{(1)}_{k}\tilde{A}^1_{k}
\! &=
-(\tilde{\cal N}^{(B)}_{k_0}\tilde{B}_{k}
+ik_\perp \,\tilde{\cal N}^{(\perp)}_{k_0}\tilde{A}^\perp_{k}), 
\\*[1.5mm]
\label{r4rg24}
((a\rho)^{-2}k_0^2+\partial_1^2-k_\perp^2+\rho^{-1}\,\partial_1)
\,\tilde{\cal N}^{(B)}_{k}\tilde{B}_{k}
\! &=0,
\end{align}
\end{subequations}
where $\partial_1=\partial/\partial \rho$ and $\partial_1^2$ means $\partial_1\partial_1$. 
Since $\tilde{A}^0_k$, $\tilde{A}^1_k$, $\tilde{A}^{\perp}_k$ and $\tilde{B}_k$ are mixed in the equations above, 
the equations have been given including the normalization constants as above.

Saving the explanation for how we have solved (\ref{r4rg2}) for later, 
we first show the mode-solutions obtained by solving (\ref{r4rg2}) in the following:
\begin{subequations}\label{ba54nwy}
\begin{align}
\label{ba54nwy1}
\bullet\quad \!
\tilde{A}^\perp_{k} 
&= \,
K_{i\alpha}(b\rho),
\\*[1.5mm]
\label{ba54nwy3}
\bullet\quad \!\! 
\tilde{A}^1_{k} 
\hspace{1.5mm} &= \,
\rho^{-1}K_{i\alpha}(b \rho),
\\*[1.5mm]
\label{ba54nwy4}
\bullet\quad \!\! 
\tilde{A}^0 _{k}
\hspace{1.5mm} &= \,
-\frac{i}{k_0\,\rho} \,\partial_1 K_{i\alpha}(b \rho)
=\frac{ib}{2k_0\,\rho}\,(K_{-1+i\alpha}(b \rho)+K_{1+i\alpha}(b \rho)),
\\*[1.5mm]
\label{ba54nwy5}
\bullet\quad \!
\tilde{B}_{k} 
\hspace{1.0mm} &= \,
ik_\perp \,K_{i\alpha}(b\rho)
= ik_\perp \,\tilde{A}^\perp_{k} ,
\end{align}
\end{subequations}
where
\begin{itemize}
\item[$\cdot$]
$\alpha \equiv k_0/a$, $b \equiv \sqrt{k_\perp^2} \,\, (\,=\sqrt{k_2^2+k_3^2})$ and $K_{i\alpha}(b\rho)$ 
is the modified Bessel function of the second kind.

\item[$\cdot$]
$\tilde{{\cal N}}^{(\perp)}_k = \tilde{{\cal N}}^{(B)}_k$ is supposed 
in the process of obtaining $\tilde{B}_{k_0,k_\perp}$.
Using (\ref{rwfyg53}), this condition leads to
\begin{eqnarray}\label{vras5g1}
{\cal N}^{(\perp)}_k \, \bm{a}^{\perp}_k = {\cal N}^{(B)}_{k}\, \bm{b}_k.
\end{eqnarray}

Here, since $\tilde{B}_{k}$ are obtained as $ik_\perp\,\tilde{A}^\perp_k$ as seen in (\ref{ba54nwy5}),
$\tilde{B}_{k}$ and $\tilde{A}^\perp_k$ are different only by the constant multiplication.
Therefore, denoting as 
$\chi_{1,k} \equiv {\cal N}^{(B)}_k\tilde{B}_k$ and 
$\chi_{2,k} \equiv {\cal N}^{(\perp)}_k\tilde{A}^\perp_k$, 
based on (\ref{e3rrjei}) and from (\ref{9beri}),
it can be written as follows:
\begin{eqnarray}\label{vhcsdo}
(\chi_{1,k},\chi_{1,k'})_{\rm KG}=(\chi_{2,k},\chi_{2,k'})_{\rm KG}= \delta(k_0-k_0')\,\delta^2(k_\perp-k_\perp').
\end{eqnarray} 
From this, it can be seen that $\chi_{1,k}=\chi_{2,k}$.
From this, it is turned out that ${\cal N}^{(B)}_k$ and ${\cal N}^{(\perp)}_k$ are related as follows: 
\begin{eqnarray}\label{vae47th}
{\cal N}^{(B)}_k=(ik_\perp)^{-1}\,{\cal N}^{(\perp)}_k. 
\end{eqnarray}

With this ${\cal N}^{(B)}_k$, it is concluded from (\ref{vras5g1}) that there is the following relation 
between the coefficients $\bm{a}^{(\perp)}_k$ and $\bm{b}_k$: $ik_\perp\,\bm{a}^{(\perp)}_k =\bm{b}_k$, 
in the classical level, which is a condition led from the supposition, $\tilde{{\cal N}}^{(\perp)}_k = \tilde{{\cal N}}^{(B)}_k$. 

\item[$\cdot$]
Next, upon solving (\ref{hwsjv}), the condition 
$\tilde{{\cal N}}^{(1)}_k = \tilde{{\cal N}}^{(0)}_k$ 
is imposed, which leads to
\begin{eqnarray}\label{vras5g2}
{\cal N}^{(1)}_k \, \bm{a}^{(1)}_k = 
{\cal N}^{(0)}_k \, \bm{a}^{(0)}_k,
\end{eqnarray}
where the decompositions (\ref{rwfyg53}) is performed. 
It turns out in (\ref{4rwerh2}) that 
${\cal N}^{(1)}_k$ and ${\cal N}^{(0)}_k$ are equivalent to each other.
Therefore, the supposition $\tilde{{\cal N}}^{(1)}_k = \tilde{{\cal N}}^{(0)}_k$ leads to  
$\bm{a}^{(1)}_k = \bm{a}^{(0)}_k$ in the classical level.

\item[$\cdot$]
From what is mentioned above, 
it can be seen that ${{\cal N}}^{(\perp)}_{k}$ and ${{\cal N}}^{(B)}_{k}$ (${{\cal N}}^{(1)}_{k}$ and ${{\cal N}}^{(0)}_{k}$) are not independent of each other;
therefore we may denote  these using a notation. 
Also, we may denote $\bm{a}^{\perp}_{k}$ and $\bm{b}_{k}$ ($\bm{a}^{1}_{k}$ and $\bm{a}^{0}_{k}$) using a notation.

However, after the canonical quantization is performed, $\bm{a}^{\perp}_{k}$ and $\bm{b}_{k}$ ($\bm{a}^{1}_{k}$ and $\bm{a}^{0}_{k}$) 
become annihilation operators in different directions. 
In this sense, $\bm{a}^{\perp}_{k}$ and $\bm{b}_{k}$ ($\bm{a}^{1}_{k}$ and $\bm{a}^{0}_{k}$) are physically distinct from each other. 
Therefore, we distinctively denote those as $\bm{a}^{\perp}_{k}$ and $\bm{b}_{k}$ ($\bm{a}^{1}_{k}$ and $\bm{a}^{0}_{k}$).

Corresponding to this, we distinctively  denote ${\cal N}^{(\perp)}_{k}$ and ${\cal N}^{(B)}_{k}$ (${\cal N}^{(1)}_{k}$ and ${\cal N}^{(0)}_{k}$) 
as ${\cal N}^{(B)}_{k}$ and ${\cal N}^{(\perp)}_{k}$ (${\cal N}^{(1)}_{k}$ and ${\cal N}^{(0)}_{k}$), as well.
\end{itemize} 
 
Now that the mode-solutions have been obtained as in (\ref{ba54nwy}), let us write $A^\mu$ in (\ref{wyea1}) as
\begin{eqnarray}\label{ds3vws}
A^\mu(\tau,\rho,x^\perp) 
\!\! &=& \!\!
\int_{-\infty}^\infty \! dk_0 \int_{-\infty}^\infty \! d^{2}k_\perp\,
{\cal N}^{(\mu)}_{k}\,\bm{a}^{(\mu)}_{k}\,\tilde{A}^\mu_{k}(\rho) \, e^{-ikx}
\nonumber\\*[1.5mm]
\!\! &=& \!\!
\Big(
\int_{-\infty}^0 \! dk_0 \,
+\int_0^\infty \! dk_0 \,
\Big)
\int_{-\infty}^\infty \! d^{2}k_\perp \,
{\cal N}^{(\mu)}_{k}\,\bm{a}^{(\mu)}_{k}\,\tilde{A}^\mu_{k}(\rho) \, e^{-ikx}
\nonumber\\*[1.5mm]
\!\! &=& \!\!
\hspace{2.0mm}
\int_0^\infty \! dk_0 \,
\int_{-\infty}^\infty \! d^{2}k_\perp \,
{\cal N}^{(\mu)}_{-k_0,-k_\perp}\,
\bm{a}^{(\mu)}_{-k_0,-k_\perp}\,
\tilde{A}^\mu_{-k_0,-k_\perp}(\rho) \, 
e^{+i({k_0 x^0-k_\perp x^{\perp}})}
\nonumber\\*[1.5mm]
&& 
\hspace{-3.5mm}
+\int_0^\infty \! dk_0 \,
\int_{-\infty}^\infty \! d^{2}k_\perp \,
{\cal N}^{(\mu)}_{+k_0,+k_\perp}\,
\bm{a}^{(\mu)}_{+k_0,+k_\perp}\,
\tilde{A}^\mu_{+k_0,+k_\perp}(\rho) \, 
e^{-i({k_0 x^0-k_\perp x^{\perp}})}.
\end{eqnarray}
In the first term in the third line above,
we will flip $k_0$ and $k_\perp$ to $-k_0$ and $-k_\perp$ respectively. 
Then, noting the following points:
\begin{itemize}

\item[$\cdot$]
There is symmetry for the $k_\perp$-direction in our system 
(actually, $\tilde{A}^\mu_{k}(\rho)$ and ${\cal N}^{(\mu)}_{k}$ are obtained independently of $k_\perp$ 
as seen in (\ref{ba54nwy}) and (\ref{4rwerh})), 
therefore, 
\begin{eqnarray}
{\cal N}^{(\mu)}_{-k_0,-k_\perp}\,
\tilde{A}^\mu_{-k_0,-k_\perp}(\rho) \, 
e^{+i({k_0 x^0-k_\perp x^{\perp}})}
\!\! &=& \!\!
{\cal N}^{(\mu)}_{-k_0,+k_\perp}\,
\tilde{A}^\mu_{-k_0,+k_\perp}(\rho) \, 
e^{+i({k_0 x^0+k_\perp x^{\perp}})},
\nonumber\\*[1.5mm]
\bm{a}^{(\mu)}_{-k_0,-k_\perp}
\!\!\! &=& \!\!
\bm{a}^{(\mu)}_{-k_0,+k_\perp}.
\nonumber
\end{eqnarray}

\item[$\cdot$]
As for the flip of $k_0$ to $-k_0$, 
all $\tilde{A}^\mu_{-k_0,+k_\perp}(\rho)$ are given by $K_{i\alpha}(b \rho)$ as seen in (\ref{ba54nwy}), 
and $K_{-i\alpha}(b \rho)=K_{+i\alpha}(b \rho)$ for  real $\alpha$, 
therefore,
$\tilde{A}^\mu_{-k_0,+k_\perp}(\rho)=\tilde{A}^\mu_{+k_0,+k_\perp}(\rho)$.
In addition, considering that $\bm{a}^{(\mu)}_{k}$ will be the annihilation operator in the quantum theory, 
we can assume $\bm{a}^{(\mu)}_{-k_0,+k_\perp}=\bm{a}^{(\mu)*}_{+k_0,+k_\perp}$.
\end{itemize}
the first term in the third line in (\ref{ds3vws}) can be equivalently rewritten as
\begin{eqnarray}\label{dwe4d}
&&
\hspace{-0.5mm}
\int_0^\infty \! dk_0 \,
\int_{-\infty}^\infty \! d^{2}k_\perp \,
{\cal N}^{(\mu)}_{-k_0,-k_\perp}\,
\bm{a}^{(\mu)}_{-k_0,-k_\perp}\,
\tilde{A}^\mu_{-k_0,-k_\perp}(\rho) \, 
e^{+i({k_0 x^0-k_\perp x^{\perp}})}
\nonumber\\*[1.5mm]
\!\! &=& \!\!
\int_{0}^\infty \! dk_0 \int_{-\infty}^\infty \! d^{2}k_\perp\,
{\cal N}^{(\mu)}_{-k_0,+k_\perp}\,\bm{a}^{(\mu)*}_{+k_0,+k_\perp}\,\tilde{A}^\mu_{+k_0,+k_\perp}(\rho) \, e^{-i({k_0 x^0-k_\perp x^{\perp}})}. 
\end{eqnarray}
$B(\tau,\rho,x^\perp)$ can be treated in the same way. 
Therefore, the expressions of the fields as the solutions are finally given as follows:
\begin{subequations}\label{w4aeea}
\begin{align}
\label{w4aeea1}
A^\mu(\tau,\rho,x^\perp) 
&=
\int_{0}^\infty \! dk_0 \int_{-\infty}^\infty \! d^{2}k_\perp\,
{\cal N}^{(\mu)}_{k}(
\bm{a}^{(\mu)}_{k}\, e^{-ikx}
+\bm{a}^{(\mu)*}_{k} \, e^{ikx})
\,\tilde{A}^\mu_{k}(\rho), 
\\*[1.5mm]
\label{w4aeea2}
B(\tau,\rho,x^\perp) &=
\int_{0}^\infty \! dk_0 \int_{-\infty}^\infty \! d^{2}k_\perp\,
{\cal N}^{(B)}_{k}(
\bm{b}_{k}\,e^{-ikx}
+\bm{b}_{k}^*\, \,e^{ikx}
)\,\tilde{B}_{k}(\rho), 
\end{align}
\end{subequations}
where we have redefined as ${\cal N}^{(\mu)}_{-k_0,+k_\perp}+{\cal N}^{(\mu)}_{-k_0,+k_\perp} \to {\cal N}^{(\mu)}_{k_0,k_\perp}$. 
Since ${\cal N}^{(\mu)}_{k}$ is the normalization constant, we may redefine like this just as the issue of the notation.
\newline

Below, we explain how we have solved (\ref{r4rg2}) and obtained (\ref{ba54nwy}).  
\begin{itemize}
\item[$\cdot$]
First, we can immediately check that $K_{i\alpha}(b\rho)$ can satisfy (\ref{r4rg23}).
Therefore, (\ref{ba54nwy1}) has been obtained.

\item[$\cdot$]
The solution of $\tilde{B}_{k}$ satisfies (\ref{r4rg24}), 
which is the same equations as (\ref{r4rg23}). 
Therefore, the solutions of $\tilde{B}_{k}$ will be proportional to $\tilde{A}^\perp_{k}$ in (\ref{ba54nwy1}), 
and its overall coefficient is the problem. 
It has been fixed based on r.h.s. of (\ref{r4rg22}):
\begin{eqnarray}\label{svdo}
\textrm{r.h.s. of (\ref{r4rg22})}=
\tilde{{\cal N}}^{(B)}_{k}\tilde{B}_{k}
+ik_\perp \,\tilde{{\cal N}}^{(\perp)}_{k}\tilde{A}^\perp_{k}=0. 
\end{eqnarray}
From this, using the solution of $\tilde{A}^\perp_{k}$ in (\ref{ba54nwy1})
and supposing $\tilde{{\cal N}}^{(B)}_{k}=\tilde{{\cal N}}^{(\perp)}_{k}$ (we discuss this in (\ref{vras5g1})), 
$\tilde{B}_{k}$ can be obtained as in (\ref{ba54nwy5}). 
  
\item[$\cdot$]
Since we have supposed the r.h.s. of (\ref{r4rg22}) as $0$ as in (\ref{svdo}), 
we can set the l.h.s. of (\ref{r4rg22}) as $0$. 
Solving this, $\tilde{A}^1_{k}$ has been obtained as in (\ref{ba54nwy3}).

\item[$\cdot$]
The r.h.s. of (\ref{r4rg25}) is essentially the same as the r.h.s. of (\ref{r4rg22}).  
It is now being taken to $0$ as in (\ref{svdo}); therefore, we can set the r.h.s. of (\ref{r4rg25}) as $0$.
From this, the following equations are obtained:
\begin{eqnarray}\label{hwsjv}
\label{hwsjv1}
\textrm{r.h.s. of (\ref{r4rg25})} 
\!\! &=\,
-ik_0 \,\tilde{{\cal N}}^{(0)}_{k}\tilde{A}^0_{k} 
+ (\partial_1+\rho^{-1})\,\tilde{{\cal N}}^{(1)}_{k}\tilde{A}^1_{k}
\nonumber\\*[1.5mm]
\!\! &=\,
-ik_0 \,\tilde{{\cal N}}^{(0)}_{k}\tilde{A}^0_{k} 
+\rho^{-1}\,\partial_1 K_{i\alpha}(b \rho)
=0,
\end{eqnarray}
where the mode-solutions of $\tilde{A}^1_{k}$ in (\ref{ba54nwy3}) and 
$\tilde{{\cal N}}^{(0)}_{k}=\tilde{{\cal N}}^{(1)}_{k}$
have been used and supposed (we discuss $\tilde{{\cal N}}^{(0)}_{k}=\tilde{{\cal N}}^{(1)}_{k}$ in (\ref{vras5g2})).
From (\ref{hwsjv1}), 
$\tilde{A}^0_{k}$ has been obtained as in (\ref{ba54nwy4}).

\item[$\cdot$]
Although (\ref{r4rg21}) has  not been used in the process above, 
it can be checked that (\ref{r4rg21}) is satisfied by the solutions $\tilde{A}^0_{k}$ and $\tilde{A}^1_{k}$ in  (\ref{ba54nwy4}) and (\ref{ba54nwy3}).

The reason for the appearance of the non-used equation is that 
the two equations (\ref{r4rg25}) and (\ref{r4rg24}) are not independent of each other, which can be seen in the description under (\ref{ebraue}). 
\end{itemize}

In the process above, the ansatz (\ref{svdo}) has been set.
However, from the perspective of the general solution, 
it is okay if only the both sides of (\ref{r4rg22}) are equivalent to each other, 
and the ansatz (\ref{svdo}) is just one situation where the equation of motion (\ref{r4rg2}) is held. 
In this sense, the solutions (\ref{ba54nwy}) represent a solution, but not the general solution.

\subsection{The normalization constants}
\label{brpnevt}

In this subsection, the normalization constants in all directions of the U(1) gauge field and $B$-field
in the RRW ((\ref{dres})) are determined from the explicit computation of the Klein-Gordon (KG) inner-product 
using the integral formulas given in Appendix\,\ref{r3g67kb}.  
The results are noted in (\ref{4rwerh}).
We begin this subsection by defining the KG inner-product in the RRW.
\newline

In general, an integral with regard to a vector on a 3D hypersurface 
in a 4D spacetime can be written as 
\begin{align}\label{d35wufa}
\int_{\Sigma} V^\mu \, d\Sigma_\mu,
\end{align}
where $V^\mu$ and $\Sigma$ mean some vector and 3D hypersurface considered, 
and $d\Sigma_\mu$ are the components of the area element on the 3D hypersurface. 

Supposing  that $ds^2$ can be denoted as $g_{00}(dx^0)^2+g_{ij}dx^idx^j$ 
($i,j=1,2,3$ and $g \equiv \det{(g_{ij})}$),
if we take the 3D hypersurface as a $x^0$-constant one,   
$V^\mu \, d\Sigma_\mu$ in (\ref{d35wufa}) is given as
\begin{eqnarray}\label{vvexbc}
V^0 d\Sigma_0 = g^{00}V_0 \, \sqrt{g_{00} g} \, d^3x. 
\end{eqnarray}

Considering our spacetime defined by (\ref{dres}), 
let us take the $\tau$-constant hypersurface in the RRW as the $x^0$-constant hypersurface considered in (\ref{vvexbc}).  
At this time, following (\ref{vvexbc}), (\ref{d35wufa}) can be written as
\begin{align}\label{dewvja}
\int_\infty^0 \, d\rho \int_{-\infty}^\infty \! d^2x^\perp  \, (a\rho)^{-1} \, V_0. 
\end{align}

Next, let us define  the  conserved current as follows:
\begin{align}\label{dberw}
J_\mu^{(f_A,g_B)}(x) \equiv i f_A^\ast(x) \overleftrightarrow{\nabla}\!_\mu \, g_B(x),
\end{align}
where $f_A^\ast \overleftrightarrow{\nabla}\!_\mu \, g_B \equiv f_A^\ast \,\nabla_\mu g_B - g_B \,\nabla_\mu f_A^\ast$, 
and $f_A$ and $g_B$ are some solutions of equations of motion.

From (\ref{dewvja}) and (\ref{dberw}), 
let us define the KG inner product 
we use as follows:
\begin{align}\label{stwgi}
(f_A,g_B)_{\rm KG} \equiv 
\int_\infty^0 \, d\rho \int_{-\infty}^\infty \! d^2x^\perp \, (a\rho)^{-1} J^{(f_A,g_B)}_0,
\end{align}
where  $(a\rho)^{-1}=\sqrt{g^{00}}$.

Using (\ref{stwgi}), we determine the normalization constants ${\cal N}^{(\mu)}_{k}$ according to the following condition:
\begin{eqnarray}\label{e3rrjei}
(
{\cal N}_{k}^{(\mu)}  \tilde{A}^\mu_{k},\,
{\cal N}_{k'}^{(\mu)} \tilde{A}^\mu_{k'}
)_{\rm KG}=
\delta(k_0-k_0')\,\delta^2(k_\perp-k_\perp'),
\end{eqnarray}
where $k$ and $k'$ in the subscripts are defined under (\ref{wyea})
($k'$ means ``$k_0',k_\perp'$''), 
and $\tilde{A}^\mu_{k}$ are given in (\ref{ba54nwy}).

From (\ref{e3rrjei}), the normalization constants defined in (\ref{rwfyg53}) can be determined as follows: 
\begin{subequations}\label{4rwerh}
\begin{align}
\label{4rwerh1}
& 
\bullet\quad \!\!  {\cal N}_{k}^{(\perp)}
= ik_{\perp}\,{\cal N}_{k}^{(B)}= 
\frac{1}{2\pi^2}
\sqrt{\frac{\sinh (\pi \alpha)}{a}},\\*[1.5mm]
\label{4rwerh2}
& 
\bullet\quad \!\!  {\cal N}_{k}^{(0)}=
{\cal N}_{k}^{(1)}= 
\frac{\alpha}{2\pi^2b} \sqrt{\frac{2\sinh (\pi\alpha)}{a}},
\end{align}
\end{subequations}
where $\alpha$ and $b$ are defined under (\ref{ba54nwy}). 
The relation between $ {\cal N}_{k}^{(\perp)}$ and ${\cal N}_{k}^{(B)}$ is the one given in (\ref{vae47th}), 
which is obtained from the condition supposed upon obtaining the solution of $\tilde{B}_{k}$.
\newline

$(
{\cal N}_{k_0}^{(\perp)} \tilde{A}^\perp_{k},\,
{\cal N}_{k'_0}^{(\perp)}\tilde{A}^\perp_{k'}
)_{\rm KG}$ 
can be written as follows:
\begin{eqnarray}\label{9beri}
&& \!\!\!
(
{\cal N}_{k}^{(\perp)} \tilde{A}^{\perp}_{k},\,
{\cal N}_{k'}^{(\perp)}\tilde{A}^{\perp}_{k'}
)_{\rm KG}
\nonumber\\*[1.5mm]
\!\!\! &=& \!\!\!
({\cal N}_{k}^{(\perp)})^2 \,\int_\infty^0 \! d\rho 
\int_{-\infty}^\infty \! \frac{d^2x^\perp}{a\rho} \,g_{\perp\perp}\,
i(A_k^{\perp*}\nabla_0 A_{k'}^\perp-A_{k'}^\perp \nabla_0 A_k^{\perp*})
\nonumber\\*[1.5mm]
\!\!\! &=& \!\!\!
({\cal N}_{k}^{(\perp)})^2 \,(2\pi)^2 \delta^2(k_\perp-k'_\perp)\,e^{i(k_0-k'_0)\tau} \, 
(\alpha+\alpha')
\int_0^\infty \! \frac{d\rho}{\rho}\, K_{i\alpha}(b\rho)K_{i\alpha'}(b\rho)
\nonumber\\*[1.5mm]
\!\!\! &=& \!\!\!
({\cal N}_{k}^{(\perp)})^2
\frac{a \, (2\pi^2)^2}{\sinh (\pi \alpha)}
\,\delta(k_0-k'_0)\,\delta^2(k_\perp-k'_\perp),
\end{eqnarray}
where in the second line, $\nabla_0 A_k^{\perp}=\partial_0 A_k^{\perp}$;
from the second to third lines, 
we put $b'=b$ based on the appearances of $\delta(k_0-k'_0)$ and $\delta^2(k_\perp-k'_\perp)$ in the equation; 
in the third line, we used (\ref{wreh11}) in Appendix\,\ref{r3g67kb}. 
From the last line, the result (\ref{4rwerh1}) can be obtained.

Next, $(
{\cal N}_{k}^{(1)}\tilde{A}^{1\ast}_{k},\,
{\cal N}_{k'}^{(1)}\tilde{A}^1_{k'}
)_{\rm KG}$ and
$
(
{\cal N}_{k}^{(0)}\tilde{A}^{0\ast}_{k},\,
{\cal N}_{k}^{(0)}\tilde{A}^0_{k'})_{\rm KG}$
can be written as 
\begin{subequations}\label{fsv45re}
\begin{align}
& \,\,
(
{\cal N}_{k_0}^{(1)}\tilde{A}^{1\ast}_{k},\,
{\cal N}_{k_0'}^{(1)}\tilde{A}^1_{k'}
)_{\rm KG}
\nonumber\\*[1.5mm]
=& \,\,
({\cal N}_{k_0}^{(1)})^2 
\int_\infty^0 \!\! d\rho 
\int_{-\infty}^\infty \! \frac{d^2x^\perp}{a\rho} \,
i\,g_{11}\,\big(
 \tilde{A}_k^{1*}\partial_0 \tilde{A}_{k'}^1
+\Gamma^1_{00}\tilde{A}_k^{1*}\tilde{A}_{k'}^0-(k' \leftrightarrow k)^*
\big),\,\,
\\*[3.0mm]
& \,\,
(
{\cal N}_{k}^{(0)}\tilde{A}^{0\ast}_{k},\,
{\cal N}_{k}^{(0)}\tilde{A}^0_{k'})_{\rm KG}
\nonumber\\*[1.5mm]
=& \,\,
({\cal N}_{k_0}^{(0)})^2 
\int_\infty^0 \!\! d\rho 
\int_{-\infty}^\infty \! \frac{d^2x^\perp}{a\rho} \,
i\,g_{00}\,\big(
\tilde{A}_k^{0*}\partial_0 \tilde{A}_{k'}^0
+\Gamma^0_{01}\tilde{A}_k^{0*}\tilde{A}_{k'}^1-(k' \leftrightarrow k)^* 
\big).\,
\end{align}
\end{subequations}
From  these, using (\ref{wreh12})-(\ref{wreh13}) in Appendix\,\ref{r3g67kb},
${\cal N}_{k}^{(0)}$ and ${\cal N}_{k}^{(1)}$ can be determined as noted in (\ref{4rwerh2}). 

\subsection{Comparison of the normalized mode-solutions with those in other studies}
\label{sbdeb}

In the subsections up until this point, 
the mode-solutions of the U(1) gauge field and $B$-field 
and those normalization constants (NC) in the Lorentz covariant gauge in the RRW of the Rindler coordinates have been obtained,   
as noted in (\ref{ba54nwy}) and (\ref{4rwerh}). 
However, those have already been addressed in other studies. 
In fact, as a result of a thorough check of the references\footnote{
The author has made the best on this point.  
}, 
\cite{Higuchi:1992td,Moretti:1996zt,Lenz:2008vw,Zhitnitsky:2010ji,Soldati:2015xma,Blommaert:2018rsf}
can be found as the studies addressing the mode-solutions of the U(1) gauge field 
in the Rindler coordinates and their NC.  
Therefore, it is considered that 
the advantages and novelties in the mode-solutions of the U(1) gauge field in the Rindler coordinates and their NC 
in our study should be discussed in comparison with 
those in the studies mentioned above, if this study provides those as an new result.
Therefore, in this subsection, the author's opinion regarding this point is given via comments on each study.
\newline

In \cite{Higuchi:1992td}, the mode-solutions are provided in its (3.8)-(3.11). 
However, how \cite{Higuchi:1992td} has solved the equations of motion would not be clear only from the description in their (3.8)-(3.11).
Also, whether or not the mode-solutions in (3.8)-(3.11) in \cite{Higuchi:1992td} can agree with ours in (\ref{ba54nwy}) is  unclear. 
This is because the notation used in the presentation of their (3.8)-(3.11) is not clear;
therefore, the method to compare their (3.8)-(3.11) with ours (\ref{ba54nwy}) is unclear.

Next, \cite{Higuchi:1992td} gives their NC in its (3.34). 
However, it is only for one direction. 
As for the consistency with our NC in our (\ref{4rwerh}), 
the NC in their (3.34) agrees with ${\cal N}_{k}^{(\perp)}$ in our (\ref{4rwerh1}). 

In this study, the calculation process to solve the equations of motion to obtain the mode-solutions 
of the U(1) gauge field  has been carefully described, and the results of the mode-solutions have been clearly provided as noted in (\ref{ba54nwy}).
Further, in this study, the NC in all directions have been obtained as noted in (\ref{4rwerh}) by performing the inner-product for each direction.
\newline

In \cite{Moretti:1996zt}, the normalized mode-solutions are provided in its (24)-(27). 
It can be seen from the description above (15) in \cite{Moretti:1996zt} that 
those (24)-(27) are provided quoting the results of \cite{Higuchi:1992td}. 
Therefore,  
corresponding to the fact that how the mode-solutions in \cite{Higuchi:1992td} have been obtained  is not clear only from  the description in \cite{Higuchi:1992td} 
as mentioned above, how (24)-(27) in \cite{Moretti:1996zt} have been obtained  is also not clear.

Turning to the NC in \cite{Moretti:1996zt}, 
the NC are included in their (24)-(27), which are all the same.
Those same NC can also be found in the description at the end of Appendix\,A in \cite{Moretti:1996zt}.
However, those same NC seem strange 
in terms of the experience of actually having solved the equations of motion and determined the NC in this study; 
that all the NC will not be the same could be expected in any situations (conversely, it seems that there is no situation where all the NC will be the same) 
if those are the NC of the mode-solution of the U(1) gauge field in the Rindler coordinates. 
Therefore, the  NC in \cite{Moretti:1996zt} seem unreliable in terms of correctness.

As for whether or not the mode-solutions in their (24)-(27)  agree with ours in (\ref{ba54nwy}), 
since those are  essentially the same type of the solutions as (3.8)-(3.11) in \cite{Higuchi:1992td},  
it is unclear for the same reason mentioned in \cite{Higuchi:1992td}, 
which is that the notation to present the mode-solutions in their (24)-(27) is unclear. 
Therefore, it is unclear how to compare those in (24)-(27) with ours in (\ref{ba54nwy}).
\newline

In \cite{Lenz:2008vw}, the $\xi$- and $\perp$-directions (the coordinates are defined in its (86)) of the U(1) gauge field 
in the Rindler coordinates in terms of the mode-solutions in the Weyl gauge are provided in its (127) and (130). 
Since the gauge taken is different  between \cite{Lenz:2008vw} and this study (the gauge in this study is Lorentz-covariant),
it might not be much meaningful to compare their mode-solutions with our mode-solutions. 
However, if checking (127) and (130) in \cite{Lenz:2008vw}, it can be seen that 
the (130), which is noted as $k_{i\frac{\omega}{a}}(z)$, agrees with ours (\ref{ba54nwy1}); 
however, the (127), which is noted as $z^2k_{i\frac{\omega}{a}}(z)$ (their $z$ is given in their (120)), does not agree with ours (\ref{ba54nwy3}). 
\cite{Lenz:2008vw} has not provided the mode-solution of the U(1) gauge field in the $\tau$-direction of the Rindler coordinates, 
but this would be because it is always 0 by the Weyl gauge \cite{Lenz:2008vw} takes. 
As for the NC,  no NC is provided in \cite{Lenz:2008vw}.
\newline

In \cite{Zhitnitsky:2010ji} and \cite{Soldati:2015xma}, 
there is no description to give the U(1) gauge field in terms of the normalized mode-solutions
(in addition, the two-dimensional Rindler coordinates are addressed in \cite{Zhitnitsky:2010ji}).
\newline

In \cite{Blommaert:2018rsf}, 
the normalized mode-solutions of the U(1) gauge field in the Rindler coordinates are provided in its (5.4). 
Since their (5.2) agrees with our normalized mode-solution in the $\perp$-direction given by (\ref{ba54nwy1}) with (\ref{4rwerh1}),
some one in their (5.4) would be the normalized mode-solution in the $\perp$-direction 
to agree with our normalized mode-solution in the $\perp$-direction given by (\ref{ba54nwy1}) with (\ref{4rwerh1}).  
However, it is unclear whether or not all mode-solutions in their (5.4) can agree with ours.
Since their notation to present their mode-solutions in (5.4) is unclear, 
comparison between their (5.4) and our (\ref{ba54nwy}) is unclear, 
as well as  the case of \cite{Higuchi:1992td} and \cite{Moretti:1996zt}.

The NC in \cite{Blommaert:2018rsf} is provided in their (5.2), 
which is the NC for the mode-solution in the $\perp$-direction
and can agree with ours in (\ref{4rwerh1}). 
However, NC other than the $\perp$-direction are not provided in \cite{Blommaert:2018rsf}.
\newline

Therefore, as the final comment in this subsection:
\begin{itemize}
\item
Regarding the mode-solutions of the U(1) gauge field in the Rindler coordinates,
all directions have been provided in \cite{Higuchi:1992td,Moretti:1996zt,Blommaert:2018rsf}. 
However, it is unclear whether or not those can agree with ours in (\ref{ba54nwy}) due to the notation used to present their mode-solutions or the difference of the gauge.

Here,   in \cite{Higuchi:1992td,Moretti:1996zt,Blommaert:2018rsf}, 
how the equations of motion have been solved
to obtain mode-solutions is unclear from their descriptions alone. 
On the other hand, in this study,
the equations of motion are clearly solved, and its process is carefully described in Sec.\,\ref{f2v4t}.  
Further, the resulting mode-solutions are noted in a fundamental manner as seen in (\ref{ba54nwy}).

This makes our mode-solutions reliable and would be an advantage compared with the mode-solutions  provided in \cite{
Higuchi:1992td,Moretti:1996zt,Blommaert:2018rsf}. 
In addition, it is no exaggeration to say that our mode-solutions constitute a new result.

\item
As for the NC,  no reference has been found in which the NC for all directions are provided. 
In such a situation, in this study, the NC in all directions have been provided in (\ref{4rwerh}) 
by clearly performing the calculation of the KG inner-product, 
which we have carefully noted in Sec.\,\ref{brpnevt} in this study. 
This is a novel point of this study.
\end{itemize}
In conclusion, this study can be said to be the first one
to properly provide the mode-solutions of the U(1) gauge field in all directions in the Rindler coordinates, including the NC.

\subsection{The canonical quantization}
\label{bywbd}
 
In this subsection, 
formulating the equal-time canonical commutation relations 
(referred to as CCR in what follows, omitting ``equal-time'') of the U(1) gauge field in the RRW ((\ref{dres})),
the commutation relations of the creation and annihilation operators are obtained in (\ref{roere}). 
Since we have obtained the mode-expanded classical solution of the U(1) gauge field in the RRW 
including the normalization constants from the straightforward analysis 
as in (\ref{ba54nwy}) and (\ref{4rwerh}),  
we can explicitly formulate the CCR. 
Then, by proceeding following the definition, 
we can obtain the result (\ref{roere}) without ambiguities or speculations.
\newline
 
Let us suppose that $A^\mu$ and $B$ satisfy the following CCR as the operators: 
\begin{subequations}\label{rntsr}
\begin{align}
\label{rntsr1}
[A^i(\tau,\rho,x^\perp),\pi_j(\tau,\rho',x'{}^\perp)] &
= \, (a\rho)^{-1} \, i\delta^i_j\delta(\rho-\rho')\delta^{2}(x^\perp-x'{}^\perp),
\\*[1.5mm]
\label{rntsr2}
[A^0(\tau,\rho,x^\perp),\pi_{0}(\tau,\rho',x'{}^\perp)] &
= \, (a\rho)^{+1} \, i\delta(\rho-\rho')\delta^{2}(x^\perp-x'{}^\perp),
\\*[1.5mm]
\label{rntsr3}
[A^\mu(\tau,\rho,x^\perp),A^\nu(\tau,\rho',x'{}^\perp)] &= 0, 
\\*[3.0mm]
\label{rntsr4}
[B(\tau,\rho,x^\perp),\pi^{(B)}(\tau,\rho',x'{}^\perp)]&
=\, (a\rho)^{+1} \, i\delta(\rho-\rho')\delta^{2}(x^\perp-x'{}^\perp), 
\\*[1.5mm]
\label{rntsr5}
[B(\tau,\rho,x^\perp),B(\tau,\rho',x'{}^\perp)]&=0,
\end{align}
\end{subequations}
where $\tau$-coordinate has been commonly taken as it plays the role of the time in the RRW, 
and $\pi_i$, $\pi_0$ and $\pi^{(B)}$ have been defined in (\ref{rr54wh}).
When $a$ is taken to $0$, (\ref{rntsr}) can agree with the CCR in the Minkowski coordinates.

Using the Christoffel symbols in (\ref{rmwg}), 
(\ref{rntsr1}), (\ref{rntsr2}) and (\ref{rntsr4}) can be written as 
\begin{subequations}\label{r6isdv}
\begin{align}
\label{r6isdv1}
[A^i(\tau,\rho,x^\perp),-\partial^0 {A}_j(\tau,\rho',x'{}^\perp)] &
= \, (a\rho)^{-1} \,i\delta^i_j\delta(\rho-\rho')\delta^{2}(x^\perp-x'{}^\perp),
\\*[1.5mm]
\label{r6isdv2}
[A^0(\tau,\rho,x^\perp),-\partial_0{A}^0(\tau,\rho',x'{}^\perp)] &
= \, (a\rho)^{+1} \,i\delta(\rho-\rho')\delta^{2}(x^\perp-x'{}^\perp), 
\\*[1.5mm]
\label{r6isdv4}
[B(\tau,\rho,x^\perp),-A^0(\tau,\rho',x'{}^\perp)]&
=\, (a\rho)^{+1} \,i\delta(\rho-\rho')\delta^{2}(x^\perp-x'{}^\perp). 
\end{align}
\end{subequations}
Using (\ref{e3ier2}) and (\ref{rntsr3}), it can be seen that (\ref{r6isdv2}) and (\ref{r6isdv4}) are equivalent to each other, 
and (\ref{r6isdv1}) and (\ref{r6isdv2}) can be written together as
\begin{eqnarray}\label{reawv}
[A^\mu(\tau,\rho,x^\perp),\partial_0 A^\nu(\tau,\rho',x'{}^\perp)] 
=- (a\rho)^{+1} g^{{\rm (M)}\mu\nu}\,i\delta(\rho-\rho')\delta^{2}(x^\perp-x'{}^\perp),
\end{eqnarray} 
where $g^{{\rm (M)}\mu\nu}={\rm diag}(+,-,-,-)$, and $(a\rho)^{+1}=\sqrt{g^{00}}^{\,-1}$.
 
When $A^\mu$ satisfies (\ref{reawv}), 
the following equation can be held for arbitrary $f_{A}$  ($A$ denotes some indices or labels) in the RRW:
\begin{eqnarray}\label{rnomt}
\!\! &&\!\! [(f_A(x), A^\mu(x))_{\rm KG},(A^\nu(y), f_B(y))_{\rm KG}]\nonumber\\*[1.5mm]
\!\! &=& \!\!
- \int \! d^3x \! \int \! d^3y 
\sqrt{g^{00}(x)}\sqrt{g^{00}(y)}
\,(
\hspace{3.2mm}
f_A^*(x)  \partial_\tau f_B(y)  \,[\partial_\tau A^\mu(x),A^\nu(y)]
\nonumber\\*[1.5mm]
&& \hspace{55.5mm}
+f_B(y) \partial_\tau f_A^*(x) \,[A^\mu(x),\partial_\tau A^\nu(y)]
) 
\nonumber\\*[1.5mm]
\!\! &=& \!\!
i\,(-g^{{\rm (M)}\mu\nu})
\int \! d^3x 
\sqrt{g^{00}(x)}
\,(
 f_A^*(x) \,\partial_\tau f_B(x) 
-f_B(x)   \,\partial_\tau f_A^*(x) 
) 
\nonumber\\*[1.5mm]
\!\! &=& \!\! - g^{{\rm (M)}\mu\nu}(f_A(x), f_B(x))_{\rm KG},
\end{eqnarray} 
where the KG inner-product defined in (\ref{stwgi}) has been used in the one above.
\newline

Here, in general, 
let us consider
some $\phi$ denoted as $\sum_i({\bm a}_i f_i+{\bm a}_i^\dagger f_i^*)$, 
where $\phi$ is a real function and $f_i$ are supposed to satisfy
$
  (f_i  ,f_j  )_{\rm KG} 
=-(f_j^*,f_i^*)_{\rm KG}
= \delta_{ij}$ and
$
  (f_i^*,f_j)_{\rm KG} 
=-(f_j,f_i^*)_{\rm KG}
=0$, 
Then, it can be checked that  
these $\phi$ and $f_i$ satisfy the following KG inner-products: 
\begin{subequations}\label{rtrsy} 
\begin{align}
  (f_i, \phi)_{\rm KG} 
=& \,
\sum_j  (f_i,{\bm a}_j f_j)_{\rm KG}
=\sum_j {\bm a}_j (f_i,f_j)_{\rm KG}
= {\bm a}_i, \\*[1.5mm]
  (\phi, f_i)_{\rm KG} 
=& \,
\sum_j   ({\bm a}_j f_j,f_i)_{\rm KG}
= \sum_j {\bm a}_j^\dagger  (f_j,f_i)_{\rm KG}
= {\bm a}_i^\dagger, \\*[1.5mm]
  (\phi, f_i^*)_{\rm KG} 
=& \,
\sum_j {\bm a}_j (f_j^*,f_i^*)_{\rm KG}
= -{\bm a}_i.
\end{align}
\end{subequations}
Then, writing (\ref{rnomt}) as 
\begin{subequations}\label{rearwsr}
\begin{align}
[(f_A,A^\mu)_{\rm KG},(A^\nu,f_B)_{\rm KG}]
=& \,
- g^{{\rm (M)}\mu\nu}\,(f_A,f_B)_{\rm KG},
\\[1.5mm]
[(f_A,A^\mu)_{\rm KG},(A^\nu,f_B^\ast)_{\rm KG}]
=& \,
- g^{{\rm (M)}\mu\nu}\,(f_A,f_B^\ast)_{\rm KG}, 
\end{align}
\end{subequations}
when $f_A$ are given by the mode-functions (\ref{ba54nwy}) normalized by (\ref{4rwerh}), 
it can be seen that the coefficients $\bm{a}^\mu_{k_0,k_\perp}$ defined in (\ref{rwfyg53}) satisfy 
the following commutation relations:
\begin{subequations}\label{roere}
\begin{align}
\label{roere1}
[\bm{a}^\mu_{k_0,k_\perp},\bm{a}^{\nu \dagger}_{k_0',k_\perp'}] 
&= 
-  g^{{\rm (M)}\mu\nu}
\delta(k_0-k_0')\,
\delta^2(k_\perp-k_\perp'), 
\\*[1.5mm]
\label{roere2}
[\bm{a}^\mu_{k_0,k_T},\bm{a}^\nu_{k_0',k_\perp'}] 
&= [\bm{a}^{\mu \dagger}_{k_0,k_\perp},\bm{a}^{\nu \dagger}_{k_0',k_\perp'}]=0. 
\end{align}
\end{subequations}
From these, it can be seen that $\bm{a}^\mu_{k_0,k_\perp}$ and $\bm{a}^{\mu \dagger}_{k_0,k_\perp}$ 
in (\ref{w4aeea}) have the physical meaning as the annihilation and creation operators, respectively, in the quantum theory.

\subsection{The polarization vector}
\label{yervd}

\begin{figure}[H]   
\vspace{0mm} 
\begin{center}
\includegraphics[clip,width=4.5cm,angle=0]{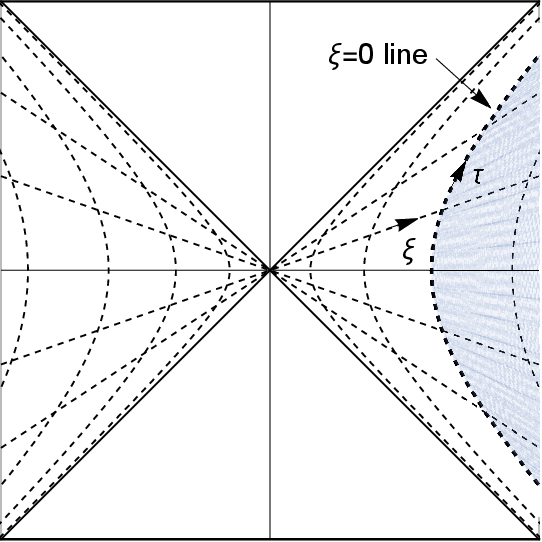} 
\end{center}
\vspace{-5.0mm}
\caption{
In this figure, supposing the bold dashed line  as the $\xi=0$ line, 
the region restricted  by (\ref{sdvlne}) is shown as the colored region. 
As for the meaning of each line, see Fig.\ref{wsdd57}.}
\label{wdet7}
\end{figure}

In the previous subsection, it was shown that 
the coefficients of the mode-solutions of the U(1) gauge field in the RRW (\ref{dres})
have the meanings as the creation and annihilation operators in the quantum theory. 
Usually, the annihilation operators of vector fields are treated 
by decomposing those into the polarization directions.
Therefore, in this section, we provide a typical polarization vector for ($S,L,\pm$)-direction 
for the annihilation operator of the U(1) gauge field in the RRW obtained in this study. 
As a result, the value of the coordinate of $\xi$ in the RRW is restricted 
as the region where the norms of the $1$-particle states in the scalar and longitudinal polarization directions are less than zero, 
as shown in Fig.\ref{wdet7}.
The origin of that restriction is discussed at the end of this subsection. 
As a result, it is concluded that those can be attributed to the non-covariance in the canonical quantization (\ref{reawv}).
\newline

Let us decompose the annihilation operator obtained in (\ref{roere}) with the polarization vector as follows:
\begin{eqnarray}\label{rvenya}
\bm{a}^\mu_{k_0,k_\perp}
= \sum_{\sigma=\pm,S,L} \varepsilon^{(\sigma)\mu}\,\bm{a}^{(\sigma)}_{k_0,k_\perp}, 
\end{eqnarray}
where $\pm$ mean positive/negative helicity directions,
and $S$ and $L$ mean the scalar and longitudinal directions.  
With this, (\ref{roere1}) can be written as
\begin{eqnarray}\label{rsytr}
\sum_{\sigma,\sigma'}\varepsilon^{(\sigma)\mu}\varepsilon^{(\sigma')\nu\ast}\,
[\bm{a}^{(\sigma)}_{k_0,k_\perp},\bm{a}^{(\sigma')\dagger}_{k'_0,k'_\perp}] 
= - g^{{\rm (M)}\mu\nu}\,\delta(k_0-k_0')\,\delta^2(k_\perp-k_\perp'),
\end{eqnarray}
where $g^{\rm (M)\mu\nu}$ is defined under (\ref{reawv}).

Now, let us introduce the matrix $\eta^{(\sigma\sigma')}$ as follows:
\begin{eqnarray}\label{rvbrek}
[\bm{a}^{(\sigma)}_{k_0,k_\perp},\bm{a}^{(\sigma')\dagger}_{k'_0,k'_\perp}] 
= \eta^{(\sigma\sigma')}\delta(k_0-k_0')\,\delta^2(k_\perp-k'_\perp), 
\end{eqnarray}
where $\eta^{(\sigma\sigma')}$ plays the role of the metric of the norms of the 1-particle states for the polarization directions.
Applying this to (\ref{rsytr}), the following relation can be obtained:
\begin{eqnarray}\label{trntr1}
g^{\rm (M)\mu\nu} 
=
- \sum_{\sigma,\sigma'}\varepsilon^{(\sigma)\mu}\,
\varepsilon^{(\sigma')\nu\ast}\,\eta^{(\sigma\sigma')}.
\end{eqnarray}
Multiplying (\ref{trntr1}) 
by $\varepsilon^{(\rho)\ast}_\mu \, \varepsilon^{(\rho')}_\nu$, 
the following relation can be obtained:
\begin{eqnarray}\label{tkoelin}
\varepsilon^{(\rho)\ast}_\mu \, \varepsilon^{(\rho')}_\nu \, g^{\rm (M)\mu\nu} 
=
-\sum_{\sigma,\sigma'} \,
(g^{\rm (R)}_{\mu\lambda} \,\varepsilon^{(\sigma)\mu}   \varepsilon^{(\rho )\lambda*})\,
(g^{\rm (R)}_{\nu\tau}    \,\varepsilon^{(\sigma')\nu *}\varepsilon^{(\rho')\tau})\,
\eta^{(\sigma\sigma')},
\end{eqnarray}
where $g^{\rm (R)\mu\nu}$ means the metric in the RRW (\ref{dres}).
~\newline

Let us consider the following $k_\mu$ as the four-dimensional momentum of the U(1) gauge field traveling in the RRW:
\begin{eqnarray}\label{reprv}
\label{reprv2}
\cdot 
\quad
\,k^\mu=(\omega,k,0,0) 
\quad
\textrm{
\!\!\! with \,\! 
$
\omega
= \frac{\vert\vec{k}\vert}{\sqrt{\vert g_{00}\vert}}
= \frac{\vert\vec{k}\vert}{\vert a  \rho \vert}
$ 
\,\! and \,\!
$
\vec{k}=(
\underbrace{\!\!k_{\,}\!\!}_{1},
\underbrace{0,0}_{\perp}
)$},
\end{eqnarray}
where $\vec{k}$ is the three-dimensional vector obtained from $k^\mu$ removing $\omega$, 
and the indices under the under-breaths mean the directions that the components the under-breaths attach to refer to.
This $k_\mu$ satisfies the condition as the momentum of the massless field: $0=k_\mu k^\mu$. 


Let us discuss how the $(S,L,\pm)$-directions of the polarization vector correspond to the directions in the Rindler coordinates.
First, from $0=k_\mu k^\mu$, 
we can see that the direction of the world-line of the U(1) gauge field  in the RRW  
(the direction of travel  in the RRW  in the Rindler coordinates) 
is either parallel or perpendicular to the Killing horizon 
(the $\xi \to -\infty$ line in  Fig.\ref{wsdd57}). 

Then, since $\tau$ plays the role of time in the RRW, 
the world-line of the U(1) gauge field in the RRW should be parametrized by $\tau$ (as noted under (\ref{dres})).
Then, the direction parallel to the Killing horizon is parametrized by $\tau$
while the direction perpendicular to the Killing horizon is $\tau$-constant; 
therefore, the direction of the world-line of the U(1) gauge field is concluded to be parallel to the Killing horizon. 

Then, since the longitudinal direction in the polarization vector is the direction of the world-line of the U(1) gauge field, 
\begin{itemize}
\item the $(S,L)$-directions of the polarization vector agree with the $(0,1)$-directions in the RRW, 
\item on the other hand, the $\pm$-directions  of the polarization vector agree with
the $\perp$-directions in the RRW up to the $O(2)$ rotation around the $L$-direction.
\end{itemize}

Now, let us assume that, 
\begin{eqnarray}\label{rrba}
\textrm{when the polarization vectors are considered, $\rho$ is taken as constant.}
\end{eqnarray}
We discuss this assumption.
The definition of $\rho$ is given in (\ref{dvyjd}), from which it can be seen that this assumption means 
to take the acceleration  as constant. 
Since we are considering the constant accelerated motion, we may impose this assumption.  

This assumption is crucial in defining the polarization vectors. 
This is because, as can be seen in (\ref{reprv}), $\omega$ depends on a coordinate $\rho$. 
Then, as can be seen later, such a $\omega$ leads to $\varepsilon^{(S)\mu}$ and $\varepsilon^{(L)\mu}$ depending on the coordinate $\rho$, 
which means that $\bm{a}^\mu_{k_0,k_\perp}$ depend on the coordinate $\rho$ as can be seen from (\ref{rvenya}).
Then, the equations of motion become unsatisfied. 
In such a situation, if $\rho$ is fixed by (\ref{rrba}), $\omega$ becomes substantially coordinate-independent.
Accordingly, $\varepsilon^{(S)\mu}$ and $\varepsilon^{(L)\mu}$ become substantially coordinate-independent, 
and the problem mentioned above does not occur. 
\newline

In the situation where $k^\mu$ is given as (\ref{reprv}), 
we will consider the following polarization vector:
\begin{subequations}\label{sceabu}
\begin{align}
\label{sceabu1}
& \bullet \quad \!\! 
\varepsilon^{(+)\mu} = -(0,0, 1, i)/\sqrt{2}, \quad
\varepsilon^{(-)\mu} = \varepsilon^{(+)\mu *}, 
\\*[1.5mm]
\label{scwehu2}
& \bullet \quad \!\!
\varepsilon^{(L)\mu}= -i k^\mu,
\\*[1.5mm]
\label{scwehu3}
& \bullet \quad \!\!
\varepsilon^{(S)\mu}= i(\omega,-\vec{k})/2\vert\vec{k}\vert^2.
\end{align}
\end{subequations}
According to the assumption (\ref{rrba}), 
this $\varepsilon^{(\sigma)\mu}$ is substantially constant for the acceleration $a$. 
Below, we mention how (\ref{sceabu}) has been obtained. 
\begin{itemize}
\item[$\cdot$]
If $\vec{k}$ is given as (\ref{reprv}), 
$\varepsilon^{(\pm)\mu}$ can  be immediately determined as (\ref{sceabu1}).  

\item[$\cdot$]
Due to the fact that the $L$-polarization direction is parallel to $k^\mu$,
$\varepsilon^{(L)\mu}$ should be a constant multiplication of $k^\mu$.
In addition, it should be able to reduce to the $L$-polarization direction in the Minkowski coordinates, 
which is $\varepsilon^{(L)\mu}_0$ in (\ref{xeadt})\footnote{
The polarization vector in the Minkowski coordinates which is referred to in the body text is the following one:
\begin{eqnarray}\label{xeadt}
\varepsilon^{(+)\mu}_0 = -(0,0, 1, i)/\sqrt{2}, \quad
\varepsilon^{(-)\mu}_0 = \varepsilon^{(+)\mu *}_0, \quad
\varepsilon^{(L)\mu}_0 = -i k^\mu, \quad
\varepsilon^{(S)\mu}_0 = i(\omega,-\vec{k})/2\vert\vec{k}\vert^2,
\end{eqnarray}
where $k^\mu$ here is $(\omega,k,0,0)$ with $\omega= \vert\vec{k}\vert$ 
($\vec{k}$ is the three-dimensional spatial vector, $(k,0,0)$). 
$\varepsilon^{(L)\mu}_0$ should be a  constant multiplication of $k^\mu$.
}, at $a=0$.
From these conditions, $\varepsilon^{(L)\mu}$ has been fixed as noted in (\ref{scwehu3}). 

Actually, since $g_{00}$ in the  RRW (\ref{dres}) reduces to $1$ at $a=0$, 
$\varepsilon^{(L)\mu}$ in (\ref{scwehu2}) can reduce to $-i(\vert k \vert, k,0,0)$ at $a=0$, 
which is $\varepsilon^{(L)\mu}_0$ in (\ref{xeadt}).

\item[$\cdot$]
Let us look at $\varepsilon^{(S)\mu}$.  
In the Minkowski case, $\varepsilon^{(S)\mu}$ is normally determined 
according to one of the equations of motion\footnote{  
In the Minkowski case, one of the equations of motion is given as $\partial_\mu A^\mu+\alpha B=0$ ($\alpha=1$). 
On the other hand, from the solution of $A^\mu$, 
$\partial_\mu A^\mu$ can be given as
\begin{eqnarray}
\partial_\mu A^\mu=
\int \frac{d^3\vec{k}}{(2\pi)^{3/2}\sqrt{2k_0}}
\sum_{\sigma=S,L\pm} \,
\big(
\bm{a}^{(\sigma)} (\vec{k})           ( -ik_\mu \varepsilon^\mu{}^{(\sigma)})     \, e^{-ikx}
+\bm{a}^{(\sigma)}{}^\dagger(\vec{k}) ( -ik_\mu \varepsilon^\mu{}^{(\sigma)}){}^* \, e^{+ikx}
\big)\big\vert_{k_0=\vert \vec{k} \vert}.
\nonumber
\end{eqnarray}
From these, in the Minkowski case, $-ik_\mu \varepsilon^\mu{}^{(\sigma)}=\delta^{\sigma S}$ is set, 
and $\varepsilon^\mu{}^{(\sigma)}$ is determined such that this can be held for the given $k_\mu$.  
}, which in this study corresponds to (\ref{ebraue4}), 
which leads to (\ref{r4rg25}).

However, in this study, $B$ is obtained not by the whole (\ref{r4rg25}), 
but by the r.h.s. of (\ref{r4rg25}) $=0$. 
Then, now  $k_\perp$ is being taken to $0$, the r.h.s. of (\ref{r4rg25}) no longer work as the relation between scalar and vector fields.

However, this means that, since there is no equation to prescribe $\varepsilon^{(S)\mu}$, we can freely take $\varepsilon^{(S)\mu}$.  
Therefore, we have set $\varepsilon^{(S)\mu}$ as noted in (\ref{scwehu2}), 
which can reduce to  $\varepsilon^{(S)\mu}_0$   (the scalar- polarization direction in the Minkowski coordinates) in (\ref{xeadt}) at $a=0$. 
\end{itemize}
~\newline

Applying (\ref{sceabu}) to (\ref{tkoelin}), we can obtain $\eta^{(\sigma\sigma')}$ as follows\footnote{
As for how to obtain (\ref{dvase}), if we assign as $\rho=\rho'=S$ in (\ref{tkoelin}), 
$(LL)$-component in the first line in  (\ref{tkoelin}) can be obtained as follows:
\begin{eqnarray}\label{rs1nt}
\varepsilon^{(S)\ast}_\mu \, \varepsilon^{(S')}_\nu \, g^{\rm (M)\mu\nu} 
=
-\sum_{\sigma,\sigma'} \,
(g^{\rm (R)}_{\mu\lambda} \,\varepsilon^{(\sigma)\mu}   \varepsilon^{(S)\lambda*})\,
(g^{\rm (R)}_{\nu\tau}    \,\varepsilon^{(\sigma')\nu *}\varepsilon^{(S)\tau})\,
\eta^{(\sigma\sigma')}
=
-\eta^{(LL)},
\end{eqnarray}
where (\ref{sceabu}) has been used. 
Performing this for other components, the first line in (\ref{dvase}) can be obtained.
The second line in (\ref{dvase}) can be obtained 
by assigning (\ref{sceabu}) to each component in the first line.
}: 
\begin{eqnarray}\label{dvase}
\eta^{(\sigma\sigma')}  
\!\! &=& \!\!
-\, \bordermatrix{
& + & - & S & L \cr
+ & 
\varepsilon^{(+)\ast}_\mu \, \varepsilon^{(+)}_\nu \, g^{\rm (M)\mu\nu} & 
\varepsilon^{(+)\ast}_\mu \, \varepsilon^{(-)}_\nu \, g^{\rm (M)\mu\nu} & 0 & 0 \cr
- & 
\varepsilon^{(-)\ast}_\mu \, \varepsilon^{(+)}_\nu \, g^{\rm (M)\mu\nu} & 
\varepsilon^{(-)\ast}_\mu \, \varepsilon^{(-)}_\nu \, g^{\rm (M)\mu\nu} & 
0 & 0 \cr
S & 
0 & 0 & 
\varepsilon^{(L)\ast}_\mu \, \varepsilon^{(L)}_\nu \, g^{\rm (M)\mu\nu} & 
\varepsilon^{(L)\ast}_\mu \, \varepsilon^{(S)}_\nu \, g^{\rm (M)\mu\nu} \cr  
L & 
0 & 0 & 
\varepsilon^{(S)\ast}_\mu \, \varepsilon^{(L)}_\nu \, g^{\rm (M)\mu\nu} & 
\varepsilon^{(S)\ast}_\mu \, \varepsilon^{(S)}_\nu \, g^{\rm (M)\mu\nu} \cr 
}
\nonumber \\*[1.5mm]
\!\! &=& \!\! \hspace{9.0mm}
\left(
\begin{array}{cccc}
1 & 0 & 0 & 0 \cr 
0 & 1 & 0 & 0 \cr 
0 & 0 & (1-g_{00})/4k_0^2 & (1+g_{00})/2  \cr
0 & 0 & (1+g_{00})/2 & (1-g_{00})\,k_0^2 
\end{array}
\right),
\end{eqnarray} 
where  
$
\varepsilon^{(\sigma)}_\mu
=
g^{\rm (R)}_{\mu\nu}\varepsilon^{(\sigma)\nu} 
$. 

Then, it can be seen from (\ref{rvbrek}) that the norms of the 1-particle states in the $S$- and $L$-directions are given as follows:
\begin{eqnarray}\label{dvnkl}
\langle 0_{\rm R}\vert 
\bm{a}^{(S)}_{k_0,k_\perp}
\bm{a}^{(S)\,\dagger}_{k_0,k_\perp} 
\vert 0_{\rm R} \rangle
= 
(1-g_{00})\,k_0^2, \quad  
\langle 0_{\rm R}\vert 
\bm{a}^{(L)}_{k_0,k_\perp}
\bm{a}^{(L)\,\dagger}_{k_0,k_\perp} 
\vert 0_{\rm R} \rangle
= 
(1-g_{00})/4k_0^2.
\end{eqnarray} 

Here, $\eta^{(SL)}=\eta^{(LS)} > 1/2$, 
which is the same situation with the Minkowski case 
in the sense that $\eta^{(SL)} $ and $ \eta^{(LS)}$ are always positive and pose no problem.   

To ensure that (\ref{dvnkl}) are less than zero, the following conditions should be satisfied:
\begin{eqnarray}\label{eael}
1-\vert g_{00}\vert  \le 0.
\end{eqnarray}
Since $g_{00}=a^2\rho^2$ and $\rho=a^{-1}e^{a\xi}$ in the RRW, 
(\ref{eael}) can be rewritten as 
\begin{eqnarray}\label{erbcne}
1 \le e^{a\xi}. 
\end{eqnarray}
From this, the region where the polarization vector (\ref{sceabu}) can be defined in the sense that 
the norms of the 1-particle states in the $S$- and $L$-directions are less than zero is restricted as follows:
\begin{eqnarray}\label{sdvlne}
\textrm{$\xi \ge 0$ for $a \ge 0$ in the RRW (\ref{dres})},
\end{eqnarray}
for $\xi \in (-\infty,\infty)$.  
The region restricted by (\ref{sdvlne}) is shown in Fig.\ref{wdet7}.
~\newline

As can be seen in (\ref{sdvlne}), (\ref{sdvlne}) are constraints which are irrelevant of the relativity.
Below, we discuss why such constraints appear.

First, (\ref{rsytr}) is the equation with regard to the quantities defined in the Rindler coordinates, 
but the metrices in its r.h.s. are not the Rindler metrices, but the Minkowski metrices; 
so, (\ref{rsytr}) is some non-covariant equation. 
Therefore, when some equation is derived based on (\ref{rsytr}), that equation is irrelevant of the relativity. 
Therefore, it can be considered that 
since $\eta^{(\sigma\sigma')}$ in (\ref{dvase}) is obtained based on (\ref{rsytr}), 
constraints irrelevant of the relativity, such as (\ref{sdvlne}), appears. 

Therefore, since the cause of the appearance of (\ref{sdvlne}) is the Minkowski metrices in (\ref{rsytr}), 
let us consider the origin of those Minkowski metrices in (\ref{rsytr}). 
Then, we can see that, 
in (\ref{reawv}), the canonical commutation relations in the Rindler coordinates are defined with the Minkowski metrices; 
once (\ref{reawv}) is given, 
(\ref{rsytr}) follows from (\ref{reawv}) by proceeding with the calculation following the definition. 
Therefore, as long as (\ref{reawv}) is considered, 
the appearance of the Minkowski metric in (\ref{rsytr}) is inevitable. 
Why the Minkowski metric appears in the equation considered in the Rindler coordinates can be considered as a general property of canonical quantization that 
the relativistic covariance is not maintained in canonical quantization.

In conclusion, the appearance of (\ref{sdvlne}) can be attributed to the general property of canonical quantization mentioned above,
and (\ref{sdvlne}) is not a problem of the classical solutions obtained in Sec.~\ref{f2v4t} and \ref{brpnevt}. 
In this sense, (\ref{sdvlne}) is no more than what specifies the region where the polarization vector (\ref{sceabu}) can be defined 
in the RRW. 

Lastly, when (\ref{sdvlne}) is not satisfied, since all four directions are positive norm states, the U(1) gauge field should be massive. 
However, we could not find any effective mass term at that time.
Yet, this would pose no problem 
because the relativistic consistency is not maintained in (\ref{sdvlne}) for the reason mentioned above.
Therefore, even if (\ref{sdvlne}) is not satisfied and all four directions become positive norm states, 
it is not necessary that the effective mass term of the U(1) gauge field must occur.

\section{The density matrix and the Unruh temperature of the U(1) gauge field in the RRW} 
\label{utempu1rc}

In this section, the density matrix of the U(1) gauge field in the RRW is obtained.
{\it The calculations performed in this section could be followed by elementary calculations}.
For each quantity in the Minkowski, LRW or RRW coordinates, 
we respectively attach 
`${\rm (M)}$', 
`${\rm (L)}$' and 
`${\rm (R)}$' 
in its superscript; 
for example, 
we denote $A^{\perp}$ 
given in the RRW 
as $A^{{\rm (R)}\perp}$ in this section.

\subsection{The expression of ${\bm a}^{{\rm (M)}\perp}_{\vec{q}}$} 
\label{usdvea}

With $V$ defined in (\ref{rebds}), 
we can denote $A^{{\rm (M)}\perp}$ as follows:
\begin{eqnarray}\label{jddliu}
A^{{\rm (M)}\perp}=
\left\{
\begin{array}{ll}
\! A^{{\rm (L)}\perp} & \!\textrm{in LRW for $V<0$,} \\[1.5mm] 
\! A^{{\rm (R)}\perp} & \!\textrm{in RRW for $V>0$,}
\end{array}
\right.
\end{eqnarray}
where the $V$-axis is presented in Fig.\ref{wsdd57}, and $A^{{\rm (M)}\perp}$ is obtained 
by $\partial_\mu x^{{\rm (L)}\perp} A^{{\rm (L)}\mu}$ 
or $\partial_\mu x^{{\rm (R)}\perp} A^{{\rm (R)}\mu}$ 
for $V<0$ or $>0$ as transformation between the contravariant vectors in the Minkowski and Rindler coordinates. 
However this is trivial, and we can write $A^{{\rm (M)}\perp}(V)$ in the following manner:
\begin{eqnarray}\label{fd6vbsi}
A^{{\rm (M)}\perp} \!\!\! &=& \!\!\!
\Theta(-V)A^{{\rm (L)}\perp} + \Theta(V)A^{{\rm (R)}\perp}, 
\end{eqnarray}  
where $\Theta(x)$ is the step function. 

We denote $A^{{\rm (L)}\perp}$ and $A^{{\rm (R)}\perp}$ as follows:
\begin{eqnarray}\label{5v3f8}
A^{{\rm (L,R)}\perp}
\equiv 
\int_0^\infty \! dk_0 
\int_{-\infty}^\infty \! d^2 k_\perp \,
\big(
\bm{a}^{{\rm (L,R)}\perp}_{k_0 k_\perp} \,
f^{{\rm (L,R)}}_{k_0 k_\perp}+\rm{c.c.}
\big),  
\end{eqnarray}
where using the results in (\ref{ba54nwy}) and (\ref{4rwerh}), 
$f^{{\rm (L,R)}}_{k_0k_\perp} $ are defined as
\begin{eqnarray}\label{qdrjb}
f^{{\rm (L,R)}}_{k_0 k_\perp} 
\equiv
{\cal N}_{k_0} K_{i\alpha}(b\vert\rho\vert)\,e^{- i k_0 \tau + i k_\perp x^\perp} 
=
\,
\sqrt{\frac{\sinh (\pi \alpha)}{a(2\pi^2)^2}} K_{i\alpha}(b\vert\rho\vert)\,
e^{- i k_0 \tau + i k_\perp x^\perp},
\end{eqnarray}
where $\rho$ is defined in (\ref{dvyjd}) and
$\alpha$ and $b$ are defined under (\ref{ba54nwy}).  

Based on the Bogoliubov transformation, 
we can write  $f^{{\rm (L,R)}}_{k_0,k_\perp}$  
in terms of modes in the Minkowski coordinate system, 
$f^{{\rm (M)}}_{q_1k_\perp} \equiv \frac{1}{\sqrt{(2\pi)^32q_0}}e^{-i(q_0t-q_1x^1-k_\perp x^\perp)}$, 
as follows:
\begin{subequations}\label{r349vd}
\begin{align}
\label{r349vd1}
\Theta(-V)\,f_{k_0 k_\perp}^{{\rm (L)}} =&
\int_{-\infty}^\infty \!
dq_1\,(
\alpha^{{\rm (L)}}_{k_0 q_1} \, 
f^{{\rm (M)}}_{q_1 k_\perp}
+\beta^{{\rm (L)}}_{k_0 q_1} \, 
f^{{\rm (M)}\ast}_{q_1 k_\perp}),
\\*[1.5mm]
\label{r349vd2}
\Theta(+V)\,f_{k_0 k_\perp}^{{\rm (R)}}=&
\int_{-\infty}^\infty \!
dq_1\,(
\alpha^{{\rm (R)}}_{k_0 q_1} f^{{\rm (M)}}_{q_1 k_\perp}
+\beta^{{\rm (R)}}_{k_0 q_1} f^{{\rm (M)}*}_{q_1 k_\perp}),
\end{align}
\end{subequations}
where $\alpha^{{\rm (L,R)}}_{k_0k_1}$ and $\beta^{{\rm (L,R)}}_{k_0k_1}$ are the Bogoliubov coefficients, 
which we will obtain in Sec.\,\ref{ugfred}. 
Applying (\ref{5v3f8}) to (\ref{fd6vbsi}), 
then with (\ref{r349vd}),  
we can give $A^{{\rm (M)}\perp}$ in (\ref{fd6vbsi}) as follows:
\begin{eqnarray}\label{fsrt}
A^{{\rm (M)}\perp} = \!
\int \! d^3\vec{q} \,\,
\Big( 
\underbrace{
\int_0^\infty \! d k_0 \,
\Big(
   \alpha_{k_0 q_1}^{{\rm (L)}}      \,  {\bm a}_{k_0 k_\perp}^{{\rm (L)}\perp}           
+ (\beta_{k_0 q_1}^{{\rm (L)}})^\ast \, ({\bm a}_{k_0 k_\perp}^{{\rm (L)}\perp})^\dagger           
+\textrm{($L \leftrightarrow R$)}
\Big)}_{\textrm{${\bm a}^{{\rm (M)}\perp}_{\vec{q}}$}}
f_{q_1 k_\perp}^{{\rm (M)}} 
+\rm{c.c.}
\Big),
\end{eqnarray} 
where $\int \! d^3\vec{q}$ denotes $\int_{-\infty}^\infty dq_1 \int_{-\infty}^\infty d^2k_\perp$. 
From (\ref{fsrt}), we can give ${\bm a}^{{\rm (M)}\perp}_{\vec{q}}$ as follows: 
\begin{equation}\label{rvsdew}
{\bm a}^{{\rm (M)}\perp}_{\vec{q}} = 
\int_{0}^\infty \! dk_0
\Big(
   \alpha_{k_0 q_1}^{{\rm (L)}}       \, {\bm a}_{k_0 k_\perp}^{{\rm (L)}\perp}           
+ ( \beta_{k_0 q_1}^{{\rm (L)}})^\ast \,({\bm a}_{k_0 k_\perp}^{{\rm (L)}\perp})^\dagger 
+\textrm{($L \leftrightarrow R$)}
\Big).
\end{equation}

\subsection{The Bogoliubov coefficients} 
\label{ugfred}

We obtain the Bogoliubov coefficients 
$\alpha_{k_0 q_1}^{{\rm (L,R)}}$ and $\beta_{k_0 q_1}^{{\rm (L,R)}}$ 
in (\ref{rvsdew}). 
For this purpose, we use the fact that 
the Bogoliubov coefficients  are independent of the coordinates and $k_\perp$. 
Then, in order to obtain the Bogoliubov coefficients, 
we can consider the KG inner product 
by the integral 
on the future Killing horizon (FKH), 
which is the $U=0$ hypersurface to be reached by $\rho \to 0$ (see Fig.\ref{wsdd57}): 
\begin{eqnarray}\label{eyjt}
(f,g)_{\textrm{KG}}^{\textrm{(FKH)}}
\equiv 
i\,{\cal V}^{-1}
\int_0^\infty \! dV \int_{-\infty}^\infty \! d^2 x^\perp
(f^\ast \partial_V g-\partial_V f^\ast g),
\end{eqnarray}
where ${\cal V} \equiv (2\pi)^{-2} \int_{-\infty}^\infty d^2x^\perp$.

The points in taking the FKH are that 
{\bf 1)} $t=x^1$  (which leads to $x^0=x^1=V/2$ based on (\ref{rebds})),
{\bf 2)} $\rho \to 0$ ($\xi \to -\infty$), and
{\bf 3)} $k_\perp=0$.
Here, when {\bf 3)} is imposed, it follows that $k_0=\pm k_1$; therefore, in what follows, we also use {\bf 3)} in the sense of $k_0=\pm k_1$.

Therefore, now $f^{{\rm (M)}}_{k_1,k_\perp}$ in (\ref{r349vd}) is given as follows:
\begin{eqnarray}\label{tst}
f^{{\rm (M)}}_{k_1,k_\perp} \vert_{{\bf 1,3)}}
= \frac{1}{\sqrt{2k_0(2\pi)^{3}}}e^{-i(k_0-k_1)V/2+ik_\perp x^\perp}\big\vert_{{\bf 3)}}.  
\end{eqnarray}
With this, 
\begin{eqnarray}\label{essrt}
(f^{{\rm (M)}}_{q_1,q_\perp}, f^{{\rm (M)}}_{k_1,k_\perp} )_{\textrm{KG}}^{\textrm{(FKH)}}
\big\vert_{{\bf 3)}}
\!\!\! &=& \!\!\!
\frac{(q_0-q_1)+(k_0-k_1)}{2\sqrt{q_0 k_0}}\,
\delta((q_0-q_1)-(k_0-k_1))
\big\vert_{{\bf 3)}}
\nonumber\\*[1.5mm]
\!\!\! &=& \!\!\!
\frac
{(\vert q_1 \vert-q_1)+(\vert k_1 \vert-k_1)}
{2\sqrt{\vert q_1 \vert \vert k_1 \vert}}\,
\delta((\vert q_1 \vert-q_1)-(\vert k_1 \vert-k_1))
\big\vert_{{\bf 3)}}.  
\end{eqnarray}

We evaluate (\ref{essrt}) by dividing it into the following cases: 
{\bf i})   $q_1 \ge 0$, $k_1 \ge 0$,
{\bf ii})  $q_1 \ge 0$, $k_1 \le 0$,
{\bf iii}) $q_1 \le 0$, $k_1 \ge 0$,
{\bf iv})  $q_1 \le 0$, $k_1 \le 0$. 
Then, in cases {\bf i}), {\bf ii}) and {\bf iii}), 
(\ref{essrt}) becomes zero, 
and from case {\bf iv}) 
the following (\ref{stukd1}) can be obtained:
\begin{subequations}\label{stukd}
\begin{align}
\label{stukd1}
(f^{{\rm (M)}}_{q_1,q_\perp}, f^{{\rm (M)}}_{k_1,k_\perp} )_{\textrm{KG}}^{\textrm{(FKH)}}\vert_{{\bf 3)}}
=& \, +\delta(\vert q_1 \vert-\vert k_1 \vert), 
\\*[1.5mm]
\label{stukd2}
(f^{{\rm (M)}}_{q_1,q_\perp}, f^{{\rm (M)}\ast}_{k_1,k_\perp} )_{\textrm{KG}}^{\textrm{(FKH)}}\vert_{{\bf 3)}}
 =& \, 0, 
\\*[1.5mm]
\label{stukd3}
(f^{{\rm (M)}\ast}_{q_1,q_\perp}, f^{{\rm (M)}\ast}_{k_1,k_\perp} )_{\textrm{KG}}^{\textrm{(FKH)}}\vert_{{\bf 3)}}
=& \, -\delta(\vert q_1 \vert-\vert k_1 \vert), 
\end{align}
\end{subequations}
where (\ref{stukd2}) and (\ref{stukd3}) have been obtained in the same way for (\ref{stukd1}).
Therefore, we can take out the Bogoliubov coefficients $\alpha_{k_0, k_1}^{{\rm (R)}}$ and $\beta_{k_0, k_1}^{{\rm (R)}}$ as follows:
\begin{subequations}
\begin{align}
\label{voier1}
(f_{q_1q_\perp}^{{\rm (M)}},\,
 f_{k_0k_\perp}^{{\rm (R)}})_{\textrm{KG}}^{\textrm{(FKH)}}\vert_{{\bf 3)}}
=&\,\,\alpha_{k_0, +k_1}^{{\rm (R)}},
\\*[1.5mm]
\label{voier2}
(-(f_{q_1q_\perp}^{{\rm (M)}})^\ast,\,
  f_{k_0k_\perp}^{{\rm (R)}})_{\textrm{KG}}^{\textrm{(FKH)}}\vert_{{\bf 3)}}
=&\,\,\beta_{k_0, +k_1}^{{\rm (R)}}. 
\end{align}
\end{subequations}

$f^{{\rm (R)}}_{k_0,k_\perp}$ is given with $K_{i \alpha}(b\rho)$ as seen in (\ref{qdrjb}).
In general, 
$K_{i \alpha}(x)
=\frac{i\pi}{2\sinh (\pi \alpha)}
\big( \frac{({x}/{2})^{ i\alpha}}{\Gamma(1+i\alpha)}
- \frac{({x}/{2})^{-i\alpha}}{\Gamma(1-i\alpha)} \big)
+{\cal O}(x^2)$. 
Therefore, by {\bf 2)}, 
$f^{{\rm (R)}}_{k_0,k_\perp}$ on the FKH can be given as
\begin{eqnarray}\label{ntrir}
f^{{\rm (R)}}_{k_0,k_\perp} \vert_{{\bf 2,3)}}
\!\!\! &=& \!\!\!
\frac{i}{2}
\frac{1}{(a \sinh (\pi \alpha))^{1/2}}
\Big(
\frac
{({b}/{2a} )^{i\alpha }e^{-ik_0 u}}
{\Gamma(1+i\alpha)}
-\frac
{({b}/{2a})^{-i\alpha}e^{-ik_0 v}}
{\Gamma(1-i\alpha)}
\Big)
\,\frac{e^{ik_\perp x^\perp} \vert_{k_\perp=0}}{2\pi}
\nonumber\\*[1.5mm]
\!\!\! &=& \!\!\!
\frac{i}{2} \,
\frac{1}{(a \sinh (\pi \alpha))^{1/2}} \,
\Big(
\frac{(b/2a)^{i\alpha}\,(-aU)^{i\alpha}}{\Gamma(1+i\alpha)}
-\frac{(b/2a)^{-i\alpha}\,(aV)^{-i\alpha}}{\Gamma(1-i\alpha)}
\Big)\, \frac{1}{2\pi}
\nonumber\\*[1.5mm]
\!\!\! &=& \!\!\!
C_+(-aU)^{i\alpha}+C_-(aV)^{-i\alpha},
\end{eqnarray} 
where  we have rewritten $u$ and $v$ to $U$ and $V$ using (\ref{rerwod}) and $C_{\pm} \equiv 
\frac{\pm i}{4\pi}(a \sinh (\pi \alpha))^{-1/2} \,
\frac{(b/2a)^{\pm i\alpha}}{\Gamma(1 \pm i\alpha)}$.
With (\ref{tst}) and (\ref{ntrir}), we can have 
\begin{eqnarray}\label{nsdtk1}
\textrm{l.h.s. of (\ref{voier1})}
=
\Big(
C_1 \, e^{-i(q_0-q_1)\frac{V}{2}} \, \frac{e^{+iq_\perp x^\perp}}{2\pi},
C_- \, (aV)^{-i \alpha} 
\Big)_{\textrm{KG}}^{\textrm{(FKH)}} \Big\vert_{{\bf 3)}}, 
\end{eqnarray} 
where $C_1 \equiv 1/\sqrt{4\pi q_0}$.
In (\ref{nsdtk1}), there is no term of $C_+$ in (\ref{ntrir}).
This is because $u \,(=\tau-\xi)$ is always diverged on the FKH 
(on FKH, $-\xi \sim \tau \sim \infty$ as can be seen from (\ref{bekwfq})), 
while $v \,(=\tau+\xi)$ takes values from $0$ to $\infty$. 
Therefore, in the integration on the FKH in terms of $\tau$ in the KG inner product (\ref{nsdtk1}),
the term of $C_+$ oscillates hard, and can be ignored. 
In the same way, we can give (\ref{voier2}) as
\begin{eqnarray}\label{nsdtk2}
\textrm{l.h.s. of (\ref{voier2})}
=
\Big(
-\Big(C_1 e^{-i(q_0-q_1)\frac{V}{2}}
\, \frac{e^{+iq_\perp x^\perp}}{2\pi}
\Big)^\ast,
C_-(aV)^{-i \alpha}
\Big)_{\textrm{KG}}^{\textrm{(FKH)}}\Big\vert_{{\bf 3)}}. 
\end{eqnarray}

We can calculate (\ref{nsdtk1}) and (\ref{nsdtk2}). 
Integrals appearing in those can be performed (using the Mathematica, etc). 
Next, $b^2=k_\perp^2$, and $k_\perp^2$ is common in the Minkowski and Rindler coordinates. 
Therefore, as the four-dimensional momentum of the massless field, 
$0=(a\rho)^2q_0^2-q_1^2-k_\perp^2=k_0^2-k_1^2-k_\perp^2$. 
From this, $b^2=k_0^2-k_1^2$, supposing $\bf 3)$.
Using this, we can finally obtain $\alpha_{k_0,+k_1}^{{\rm (R)}}$ as follows:
\begin{eqnarray}\label{rsdv}
\bullet \quad
\alpha_{k_0,+k_1}^{{\rm (R)}}=
-e^{\pi \alpha} \beta_{k_0,+k_1}^{{\rm (R)}}=
\frac{1}{\sqrt{2\pi k_0 a}}
\frac
{\big(\frac{k_0-k_1}{k_0+k_1}\big)^{i\alpha/2}}
{(1-e^{-2\pi \alpha})^{1/2}}.
\end{eqnarray}
We can get $\alpha_{k_0,k_1}^{{\rm (L)}}$ and $\beta_{k_0,k_1}^{{\rm (L)}}$ 
from the following links:
\begin{eqnarray}\label{erbenr}
\bullet \quad
\alpha_{k_0,-k_1}^{{\rm (L)}}= \alpha_{k_0,+k_1}^{{\rm (R)}}, \quad
\beta_{k_0,-k_1}^{{\rm (L)}}= \beta_{k_0,+k_1}^{{\rm (R)}}. 
\end{eqnarray}

\subsection{The Minkowski ground state in the U(1) gauge field} 
\label{eppwte}

We can perform the following computations as follows:
\begin{subequations}\label{sdv509}
\begin{align}
\label{sdv5091}
\int^\infty_{-\infty} \! dq_1 \, 
\alpha_{p_0q_1}^{{\rm (R)}}\,
(\alpha_{k_0q_1}^{{\rm (R)}})^\ast \big\vert_{{\bf 3)}}
=& \, e^{\pi q/a}  
\int^\infty_{-\infty} \! dk_1 \, 
\alpha_{p_0k_0}^{{\rm (R)}}\,
(\beta_{q_0k_0}^{{\rm (R)}})^\ast \big\vert_{{\bf 3)}}
\nonumber\\*[1.5mm]
=& \, e^{2\pi q/a} 
\int^\infty_{-\infty} \! dk_1 \, 
 \beta_{p_0k_0}^{{\rm (R)}} \, 
(\beta_{q_0k_0}^{{\rm (R)}})^\ast \big\vert_{{\bf 3)}}
=
\frac{\delta(p_0-k_0)}{1-e^{-2\pi p_0/a}},
\\*[1.5mm]
\label{sdv5093}
      \int^\infty_{-\infty} \! dk_1 \, \alpha_{p_0k_0}^{{\rm (R)}}\,\alpha_{q_0k_0}^{{\rm (R)}} \big\vert_{{\bf 3)}}
=& \, \int^\infty_{-\infty} \! dk_1 \, \alpha_{p_0k_0}^{{\rm (R)}}\,\beta_{q_0k_0}^{{\rm (R)}} \big\vert_{{\bf 3)}}
=0,
\end{align}
\end{subequations}
where the integrals appearing in the computations can be performed and become delta-functions by 
$
\frac{1}{2\pi a}
\int_{-\infty}^\infty dq_3 
\frac{1}{q_0}
(\frac{q_0-q_1}{q_0+q_1})^{i (p_0-k_0)/2a}
=
\frac{1}{2\pi a}
\int_{-\infty}^\infty d\theta \,
e^{i \theta(p_0-k_0)/{a}}
=\delta(p_0-k_0)
$ 
$\big(
\theta \equiv \frac{1}{2}\log (\frac{k_0+k_1}{k_0-k_1})
\big\vert_{{\bf 3)}}
= \frac{1}{2}\log (\frac{\sqrt{k_1^2+k_\perp^2}+k_1}{\sqrt{k_1^2+k_\perp^2}-k_1})
\big\vert_{{\bf 3)}}
\big)$.

Using (\ref{sdv509}) and ${\bm a}^{{\rm (M)}\perp}_{\vec{q}}$ given in (\ref{rvsdew}), 
we can obtain the following result:
\begin{subequations}\label{srtts}
\begin{align}
\int^\infty_{-\infty} \! dq_1 \,
\alpha_{p_0 q_1}^{{\rm (R,L)}} {\bm a}^{{\rm (M)}\perp}_{\vec{q}}  
= 
\frac{{\bm a}^{{\rm (L,R)}\perp}_{p_0 k_\perp}-e^{-\pi p_0/a}({\bm a}^{{\rm (R,L)}\perp}_{p_0 k_\perp})^\dagger}
{1-e^{-2\pi {p_0}/a}}. 
\end{align}
\end{subequations}
Therefore, 
from the identity 
${\bm a}^{{\rm (M)} \perp}_{\vec{k}\perp} \vert 0_{\rm M} \rangle =0$, 
we can obtain the following identity:
\begin{eqnarray}\label{beqver}
({\bm a}^{{\rm (L,R)}\perp}_{p_0 k_\perp}-e^{-\pi p_0/a}({\bm a}^{{\rm (R,L)}\perp}_{p_0 k_\perp})^\dagger)
|0_{\rm M}\rangle = 0.
\end{eqnarray} 
Then, by 
$
 ({\bm a}^{{\rm (L)}\perp}_{p_0 k_\perp})^\dagger \times (\ref{beqver})
-({\bm a}^{{\rm (R)}\perp}_{p_0 k_\perp})^\dagger \times (\ref{beqver})
$, we can get the following relation:
\begin{eqnarray}\label{ssydv}
({\bm a}^{{\rm (L)}\perp}_{p_0 k_\perp})^\dagger 
{\bm a}^{{\rm (L)}\perp}_{p_0 k_\perp} \,
|0_{\rm M}\rangle
=({\bm a}^{{\rm (R)}\perp}_{p_0 k_\perp})^\dagger 
{\bm a}^{{\rm (R)}\perp}_{p_0 k_\perp} \,
|0_{\rm M}\rangle. 
\end{eqnarray}  
(\ref{ssydv}) means that 
the Minkowski ground state constitutes a linear combination of the outer product of the left and right Rindler states 
which have the same excited numbers as each other, namely,
\begin{eqnarray}\label{b7user}
\vert 0_{\rm M} \rangle 
\equiv
\sum_{n_1,n_2, \cdots=0}^\infty
\! {\cal K}_{n_1,n_2, \cdots} \,
\Big[
\prod_{i=1}^\infty  
\Big(
({\bm a}^{{\rm (L)}\perp}_{\omega_i})^\dagger 
({\bm a}^{{\rm (R)}\perp}_{\omega_i})^\dagger
\Big)^{n_i}
\Big]\,
\vert 0_{\rm L} \rangle \otimes \vert 0_{\rm R} \rangle,
\end{eqnarray}
where $\omega_i$ is an abbreviated expression of $p_0^{(i)} k_\perp^{(i)}$, 
and ${\cal K}_{n_1,n_2, \cdots}$ are coefficients of the linear combination.

Let us denote the r.h.s. of (\ref{b7user}) as $\Omega \, \vert 0_{\rm L} \rangle \otimes \vert 0_{\rm R} \rangle$, 
where
\begin{eqnarray}\label{bsr3q}
\Omega 
=
\prod_{i=1}^\infty \,
\Big[
\sum_{n_i=0}^\infty \,
{\cal J}_{n_i}  \,
\Big(
({\bm a}^{{\rm (L)}\perp}_{\omega_i})^\dagger 
({\bm a}^{{\rm (R)}\perp}_{\omega_i})^\dagger
\Big)^{n_i}
\Big].
\end{eqnarray}
${\cal J}_{n_i}$ can be recursively obtained according to the condition (\ref{beqver})\footnote{
For some $i$, 
$
\vert 0_{\rm M} \rangle \sim 
  \big(
  {\cal J}_0 
+ {\cal J}_1 \,
  {\bm a}^{{\rm (L)}\dagger}_i
  {\bm a}^{{\rm (R)}\dagger}_i    
+ {\cal J}_2 \,
  ({\bm a}^{{\rm (L)}\dagger}_i 
   {\bm a}^{{\rm (R)}\dagger}_i)^2 +\cdots
\big)
\,\vert 0_{\rm L} \rangle \otimes \vert 0_{\rm R} \rangle
$. 
From the $\vert 0_{\rm M} \rangle$ multiplied by ${\bm a}^{{\rm (R)}}_{i}$ 
and the $\vert 0_{\rm M} \rangle$ multiplied by ${\bm a}^{{\rm (L)}\dagger}_{i} e^{-\pi p}$,
\begin{subequations}\label{ua344}
\begin{align}
\label{ua3441}
&
\big( 
{\cal J}_1\,  {\bm a}^{{\rm (L)}\dagger}_i
+ 2{\cal J}_2 \, {\bm a}^{{\rm (R)}\dagger}_i ({\bm a}^{{\rm (L)}\dagger}_i)^2 +\cdots
\big) 
\,\vert 0_{\rm L} \rangle \otimes \vert 0_{\rm R} \rangle
,
\\*[1.5mm]
\label{ua3442}
&
\big( 
{\cal J}_0 {\bm a}^{{\rm (L)}\dagger}_{i}
+  {\cal J}_1 \, {\bm a}^{{\rm (R)}\dagger}_i ({\bm a}^{{\rm (L)}\dagger}_i)^2 
+ 2{\cal J}_2 \, ({\bm a}^{{\rm (R)}\dagger}_i)^2 ({\bm a}^{{\rm (L)}\dagger}_i)^3
+\cdots
\big) 
\, e^{-\pi p}\, \vert 0_{\rm L} 
\,\rangle \otimes \vert 0_{\rm R} \rangle,
\end{align}
\end{subequations}
where roughly $[{\bm a}^{\rm (R)}_{i},{\bm a}^{\rm (R)\dagger}_{i}]\sim 1$. 
Using (\ref{beqver}), $
\textrm{(\ref{ua3441})}-\textrm{(\ref{ua3442})}
=\big(
{\bm a}^{{\rm (R)}}_{i} -{\bm a}^{{\rm (L)}\dagger}_{i} e^{-\pi p} 
\big) 
\,\vert 0_{\rm L} \rangle \otimes \vert 0_{\rm R} \rangle=0$, while
\begin{eqnarray}
\textrm{(\ref{ua3441})}-\textrm{(\ref{ua3442})}=
({\cal J}_1-{\cal J}_0\, e^{-\pi p}){\bm a}^{{\rm (L)}\dagger}_i
+(2{\cal J}_2-{\cal J}_1\, e^{-\pi p}) {\bm a}^{{\rm (R)}\dagger}_i ({\bm a}^{{\rm (L)}\dagger}_i)^2
+\cdots. 
\end{eqnarray}
Therefore, ${\cal J}_1={\cal J}_0\, e^{-\pi p}$, ${\cal J}_2={\cal J}_0\, e^{-2\pi p}/2$, $\cdots$, and we can get (\ref{wrbsdv}).
}, 
from which we can  give the Minkowski ground state as follows:
\begin{eqnarray}\label{wrbsdv}
\vert 0_{\rm M} \rangle 
\!\!\! &=& \!\!\!
\prod_{i=1}^\infty \,
\Big[{\cal J}_{0_i}
\sum_{n_i=0}^\infty
\frac{e^{-\pi/a \cdot n_i p_0^{(i)}}}{n_i!} 
\Big(
({\bm a}^{{\rm (L)}\perp}_{\omega_i})^\dagger 
({\bm a}^{{\rm (R)}\perp}_{\omega_i})^\dagger
\Big)^{n_i}
\Big]\,
\vert 0_{\rm L} \rangle \otimes \vert 0_{\rm R} \rangle,
\end{eqnarray}
where ${\cal J}_{0_i}$ are constants of the first terms in each series of ${\cal J}_{n_i}$ (see footnote).

\subsection{The density matrix and the Unruh temperature of the U(1) gauge field in the RRW} 
\label{yedvsh}

In (\ref{wrbsdv}), putting all ${\cal J}_{0_i}$ as $1$, 
we can obtain the density matrix in the RRW as\footnote{
Using the orthogonal vectors in the state space of the LRW $\vert n_{i, \rm L} \rangle 
\equiv 
\frac{\big(({\bm a}^{{\rm (L)}\perp}_{\omega_i})^\dagger\big)^{n_i}}{\sqrt{n_i!}}
\vert 0_{\rm L} \rangle$,
\begin{eqnarray}\label{waejt}
{\rm Tr}_{\rm L}
\big[\vert 0_{\rm M} \rangle \langle 0_{\rm M} \vert\big] 
\! = \!
\prod_i 
\big[\sum_{n_i}
\langle n_{i, \rm L} \vert 0_{\rm M} \rangle \langle 0_{\rm M}  \vert n_{i, \rm L} \rangle
\big] 
\! = \!
\prod_i \,
\big[
\sum_{n_i}
e^{-2\pi/a \cdot n_i p_0^{(i)}}
\,
\vert n_{i, \rm R} \rangle
\langle n_{i, \rm R} \vert
\big]
= \textrm{(\ref{wydrb})}.
\end{eqnarray}
}
\begin{eqnarray}\label{wydrb}
\hat{\rho}_{\rm R} 
=
{\rm Tr}_{\rm L}
\big[\vert 0_{\rm M} \rangle \langle 0_{\rm M} \vert\big] 
=
\frac
{e^{-\beta\,\hat{H}_{\rm R}}}
{{\rm Tr}_{\rm R}(e^{-\beta\,\hat{H}_{\rm R}})},
\end{eqnarray}
where $\beta=2\pi/a$.
From this we can see that the Minkowski ground state in the RRW is the density matrix 
in temperature $\beta^{-1}$. Therefore, we can conclude that 
the U(1) gauge field in a constant accelerated system with the acceleration $a$ 
will feel the Unruh temperature $\beta^{-1}=a/2\pi$, which is given (\ref{hertrt}), as well.

\section{Summary} 
\label{Summary}

In this section, this study is summarized.
Obtaining the Lagrangian of the U(1) gauge field by the Lorentz-covariant gauge
in the right Rindler-wedge (RRW) of the Rindler coordinates as seen in Sec.\,\ref{f367vo}, 
we have obtained the mode-solutions by straightforwardly solving the equations of motion 
obtained from that gauge-fixed Lagrangian, as seen in Sec.\,\ref{f2v4t}.

Next, calculating the integrals as seen in Appendix\,\ref{r3g67kb} and using the results, 
we have determined the normalization constants (NC) in all directions of those mode-solutions in the RRW
from the explicit calculation based on the KG inner-product, 
as seen in Sec.\,\ref{brpnevt}.

Following that, 
based on those normalized mode-expanded classical solutions of the U(1) gauge field obtained in such a clear manner, 
and formulating the equal-time canonical commutation relations of the U(1) gauge field, 
we have given the commutation relations of the creation and annihilation operators in the RRW, as seen in Sec.\,\ref{bywbd}.
\newline

The canonical quantization (CQ) of the scalar and spinor field 
in the Rindler coordinates has already been conducted  in \cite{Higuchi:2017gcd} and \cite{Soffel:1980kx,Ueda:2021nln}, respectively;
however, the CQ of the gauge fields in the Rindler coordinates has not been properly performed yet 
(the reason for this has been mentioned in Sec.\,\ref{sec:int}).
Here, it is important to stress that the NC in other previous studies are considered 
to have been provided incorrectly. 
Considering this situation, 
this study would be the first to obtain the mode-solutions, 
including the NC in all directions, in the RRW of the Rindler coordinates
by solving the equations of motion and by explicitly calculating the KG inner-products. 
The points mentioned here have been discussed in Sec.\,\ref{sbdeb} 
via the checking of the mode-solutions and NC provided in other studies. 

The mode-solutions obtained in this study are not general solutions, as mentioned at the end of Sec.\,\ref{f2v4t}. 
However, 
considering that 
our mode-solutions and those NC have been obtained from explicit calculations unlike those in other studies, 
where the NC in other studies are considered to be incorrect (these points have been discussed in Sec.\,\ref{sbdeb}),
it is believed that the normalized mode-solutions obtained in this study provide a significant contribution to the field.
\newline

After performing the canonical quantization, we have provided a typical polarization vector for the ($S,L,\pm$)-direction in Sec.\,\ref{yervd}.  
As a result, it has been found that a constraint for the Rindler coordinate arises when  considering that typical polarization vector. 
We have discussed its origin at the end of Sec.\,\ref{yervd}.  
Then, in Sec.\,\ref{utempu1rc},
we have obtained the density matrix of the U(1) gauge field in the RRW, 
and shown that the U(1) gauge field will feel the Unruh temperature.
\newline

As for the future development of this study, we would first consider extending the U(1) gauge field in this study to the non-Abelian gauge field. 
However, this would not be feasible, because, if the gauge field considered is non-Abelian, 
the interaction terms between the gauge fields are involved in the equations of motion of the gauge field. 
As a result, due to the existence of those interaction terms in the equations of motion,
obtaining the analytical solution becomes infeasible even in the Minkowski coordinates.

There is also an interesting study \cite{Hotta:2016qtv}, 
in which the gravitational charges in the neighborhood of the Killing horizon in the Rindler coordinates are defined. 
Therefore, based on this study, we can expect that there is some asymptotic symmetry in the Rindler coordinates. 
This would lead to some holographic dual CFT to the Rindler coordinates, 
which is also very interesting, in addition to the AdS/CFT and Kerr/CFT correspondences.
One of the issues to be concretely addressed in the analysis of the asymptotic symmetry is to check the equality between 
the Ward identities and the scattering amplitudes including the soft-particles in the asymptotically flat spacetime~\cite{Strominger:2017zoo,Raclariu:2021zjz}.
Since this is a scattering amplitude analysis,
the creation and annihilation operators of the gauge or gravitational field are explicitly used. 
Therefore, in this future direction, the normalized mode-solutions of the U(1) gauge field obtained in this study would be useful.
However, in this analysis, the fact that the spacetime  can be treated as a causal diamond is essential 
upon addressing the incoming and outgoing particles for the system. 
In the case of asymptotically flat spacetime,  
this is possible by using the Penrose coordinates; 
however, the Rindler coordinates cannot be rewritten into the closed system in principle.
In this sense, in the study of the asymptotic symmetry in the Rindler coordinates 
based on the analysis to check the equality between 
the Ward identities and the scattering amplitudes including the soft-particles, 
there would arise new problem  which does not exist
in the case of asymptotically flat spacetime.

Next, we have obtained the density matrix of the U(1) gauge field in Sec.\,\ref{yedvsh}. 
Then, in terms of the Bekenstein-Hawking entropy, 
it is interesting to analyze the entanglement entropy (EE) 
between the U(1) gauge fields in the LRW and RRW 
based on the von Neumann entropy 
(using the replica trick and heat kernel regularization).
The EE in the causally separated spaces has already been analyzed in various models in many studies, 
and its leading contribution and the corrections to that 
have been discussed in terms of black hole issues.
In the analysis of EE of the U(1) gauge field, \cite{Kabat:1995eq} would be helpful. 
 
As a phenomenological future direction
in which the creation and annihilation operators of the U(1) gauge field 
in the Rindler coordinates are used, 
we can consider studies on the Unruh effect in photon antibunching. 
In an analysis of this future study, 
\cite{Giovannini:2010xg} would be a helpful reference. 
Currently, many ways to detect the Unruh effect have been investigated as cited in Sec.\,\ref{sec:int}, 
and this future study could be interesting as one of those new ways.
~\newline
~\newline
{\bf Acknowledgment}~~This work was conducted with the Mathematica of the Yukawa Institute Computer Facility.
  
\appendix

\section{Replacement of the Coulomb gauge with the Lorentz-covariant gauge in the path-integral in the RRW}
\label{buobhs}

Let us replace the gauge fixing conditions in (\ref{tedrha}) with the Lorentz-covariant gauge.  
For this purpose, 
considering $f={\cal C}$ 
as a Lorentz-covariant gauge condition 
(where $f=\nabla_\mu A^\mu$ and ${\cal C}$ is some real function as given under (\ref{ebtwf})), 
let us consider the following Faddeev-Popov determinant:
\begin{eqnarray}\label{sbrtv}
\triangle_f[A]^{-1} \equiv 
\int {\cal D}U \! \prod_{x \in {\rm RRW}} \delta(f[A^U]-{\cal C}(x)),
\end{eqnarray}
where $A^U$ denotes a gauge transformed $A^\mu$ 
such as  $A^\mu+\nabla^\mu U$, 
and
\begin{eqnarray}\label{sbrtv2}
\triangle_f[A] = {\rm Det}\big[\nabla_\mu \nabla^\mu  \,\delta^4(x-y)\big],
\end{eqnarray}
where note that $\triangle_f[A]$ does not include the gauge fields if the gauge field is the U(1), 
which leads to the situation that the gauge and ghost fields do not couple each other (as can be seen from (\ref{edsstr})).

Then, inserting the unity obtained from (\ref{sbrtv}) into (\ref{tedrha}), 
and exploiting the gauge invariance,
we can write (\ref{tedrha}) as follows:
\begin{eqnarray}\label{erbi75}
\textrm{(\ref{tedrha})}
\!\!\! &=& \!\!\!
\int \! {\cal D}\!A^k \,
\Big[
\int \! {\cal D}U \! \prod_{x \in {\rm RRW}} 
\big[\delta(\chi^{(2)})\big] \cdot
\prod_{\tau} \textrm{Det}\big[M_c(x,y)\big]
\Big]
\,
\triangle_f[A]
\prod_{x \in {\rm RRW}}
\big[\delta(f[A]-{\cal C})\big] 
\nonumber\\*[1.5mm]
&& \!\!\!\! \times\,   
\exp \big[i\int_{\rm RRW} d^4x \,\sqrt{-g}\,{\cal L}_{\rm U(1)} \big].
\end{eqnarray}
Here, 
\begin{eqnarray}
\label{sbrtv1}
\int \! {\cal D}U \! \prod_{x \in {\rm RRW}} \big[\delta(\chi^{(2)}) 
\big]  
&= \Big(\prod_{\tau} {\rm Det}
\big[\nabla_k \nabla^k \,\delta^3(x-y)\big]
\Big)^{-1},
\end{eqnarray}
which is the inverse of $\prod_{\tau} \textrm{Det}[M_c(x,y)]$ 
($M_c(x,y)$ is defined in (\ref{neic1})); therefore,
\begin{eqnarray}\label{4reuo4}
\textrm{(\ref{erbi75})}
\!\!\! &=& \!\!\!
\int \! {\cal D}\!A^k \,
\triangle_f[A]
\prod_{x \in {\rm RRW}}
\big[\delta(f[A]-{\cal C})\big] \cdot   
\exp \big[i \int_{\rm RRW} d^4x \,\sqrt{-g} \,{\cal L}_{\rm U(1)} \big].
\end{eqnarray}

Rewriting $\triangle_f[A]$ in (\ref{sbrtv2}) 
with the ghost and anti-ghost fields $c$ and $\bar{c}$ as 
\begin{eqnarray}\label{edsstr}
\triangle_f[A]=
\int \! {\cal D}c\,{\cal D}\bar{c} \, 
\exp \, [ \, 
i \! \int_{\rm RRW} \! d^4x \, \sqrt{-g} \, 
i \,\bar{c} \, \nabla_\mu \nabla^\mu  c
],
\end{eqnarray}
we incorporate $\triangle_f[A]$ into the Lagrangian.
Then, inserting the following unity,
\begin{eqnarray}\label{bb54va}
1= 
\int \! {\cal D}B\, {\cal D}{\cal C}
\exp \big[ \int_{\rm RRW} \! d^4x \, \sqrt{-g} \, (B\,{\cal C}+ B^2/2) \big], 
\end{eqnarray}
we integrate out ${\cal C}$. 
As a result, we can obtain the Lagrangian in (\ref{etsiph}). 

\section{The integral formulas used to determine the normalization constants}
\label{r3g67kb}

In this Appendix, we give the integral formulas 
used in Sec.\,\ref{brpnevt} 
to determine the normalization constants of the mode-solutions (\ref{wyea}) 
of the U(1) gauge field in the RRW in the Rindler coordinates. 
Those are essential in the computation of the KG inner-product between those mode-solutions. 
We have obtained those integral formulas in this study to perform those KG inner-product. 
\newline 

First, we note the integrals and the results of those in the following:
\begin{subequations}\label{wreh1}
\begin{align}
\label{wreh11}
\int_0^\infty \! dx \, x^{-1}K_{iu}(x)K_{iv}(x)
=& \,
\frac{1}{2u}\frac{\pi^2}{\sinh(\pi u)}\delta(u-v),\\*[1.5mm]
\label{wreh12}
\int_0^\infty \! dx \, x^{-3}K_{iu}(x)K_{iv}(x)
=& \,
\frac{1}{4u(1+u^2)}\frac{\pi^2}{\sinh(\pi u)}\delta(u-v),\\*[1.5mm]
\label{wreh14}
\int_0^\infty \! dx \, x^{-2} K_{iu}(x)(K_{-1+iv}(x)+K_{1+iv}(x))
=& \,
-\frac{1}{2u(1+u^2)}\frac{\pi^2}{\sinh(\pi u)}\delta(u-v),\\*[1.5mm]
\label{wreh13}
\int_0^\infty \! dx \, x^{-1}
(K_{-1+iu}(x)+K_{1+iu}(x))
(u \leftrightarrow v)
=& \,
-\frac{u}{1+u^2}
\frac{\pi^2}{\sinh(\pi u)}\delta(u-v).
\end{align}
\end{subequations}
The definition range of (\ref{wreh11}) is 
out of the integral formula in \cite{Gradshteyn:1943cpj} (4 of 6.576),  
and the remaining three are not included in \cite{Gradshteyn:1943cpj}.
Therefore, we must obtain these by ourselves. 
In what follows, how we have obtained (\ref{wreh11}) is shown.
(\ref{wreh12})-(\ref{wreh13}) can be obtained in the same way. 
\newline

Using the formula in \cite{Gradshteyn:1943cpj} (4 of 6.576), 
we will give the l.h.s. of (\ref{wreh11}) replacing $x^{-1}$ with $x^{-(1-\omega)}$ ($\omega$ is finally taken to $0$)
as follows:
\begin{eqnarray}\label{vsdvbs1}
&& \!\!
\int_0^\infty \! dx \, x^{-(1-\omega)}K_{iu}(x)K_{iv}(x)
\nonumber\\*
\!\! &=& \!\!
\frac{2^{-2-(1-\omega)}}{\Gamma(1-(1-\omega))}
\Gamma(\frac{1-(1-\omega)+i(u+v)}{2})
\Gamma(\frac{1-(1-\omega)+i(-u+v)}{2})
\nonumber\\*
&& \!\! 
\Gamma(\frac{1-(1-\omega)+i(u-v)}{2})
\Gamma(\frac{1-(1-\omega)-i(u+v)}{2}).
\end{eqnarray}

We can set all $\omega$ as zero except for that in $\Gamma(\omega)$.  
As a result, (\ref{vsdvbs1}) reduces as follows: 
\begin{eqnarray}\label{vhscu}
\textrm{(\ref{vsdvbs1})}
\!\! &=& \!\!
\frac{2^{-3}}{\Gamma(\omega)} 
\, \Gamma(\frac{i(u+v)}{2}) 
\, \Gamma(\frac{i(-u+v)}{2})
\, \Gamma(\frac{i(u-v)}{2})
\, \Gamma(\frac{-i(u+v)}{2})
\nonumber\\*[1.5mm]
\!\! &=& \!\!
2^{-3}\,\omega
\, \frac{\pi}{\frac{u-v}{2}\sinh(\pi\frac{u-v}{2})}
\, \frac{\pi}{\frac{u+v}{2}\sinh(\pi \frac{u+v}{2})}+{\cal O}(\omega^2),
\end{eqnarray}
where $1/\Gamma(\omega)=\omega+{\cal O}(\omega^2)$.

Since $\omega$ is taken to $0$,
(\ref{vhscu}) vanishes  for the case $u \not=v$; namely,
\begin{eqnarray}\label{rnast}
\textrm{(\ref{vhscu})}=0 \quad \textrm{for $u \not=v$}.
\end{eqnarray}
On the other hand, in the case $u =v$, 
we suppose that $u-v=i \varepsilon$ ($\varepsilon$ is finally taken to $0$).
Then, we suppose $\varepsilon$ as $\varepsilon=\omega$ as the same infinitesimal quantity.
As a result, (\ref{vhscu}) can be calculated as
\begin{eqnarray}\label{vsdvbs3}
\textrm{(\ref{vhscu})}
= 
\frac{-i}{\pi}
\frac{1}{i\varepsilon}
\frac{\pi^2}{2u \sinh(\pi u)}+{\cal O}(\varepsilon^2),
\end{eqnarray}
where the expansion around $\varepsilon=0$ has been performed making use of the fact that $\varepsilon$ is finally taken to $0$, 
and $u$ has been set as $v$ at the stage of (\ref{vsdvbs3}) by considering the fact that $\delta(u-v)$ finally appears. 

Then, there is a general relation held as the relation of the integrand: 
$\lim_{\lambda \to 0}\frac{1}{x-i\lambda}={\rm p.v.}\frac{1}{x}+\pi i \,\delta(x)$. 
From this relation, $\delta(x)$ can be written as
$\delta(x)
= \frac{1}{\pi}{\rm Im}\big[\lim_{\lambda \to 0}\frac{1}{x-i\lambda}\big]
= \frac{-i}{\pi}\lim_{\lambda \to 0}\frac{1}{x-i\lambda}
$. 
Using this expression, (\ref{vsdvbs3}) can be written as follows:
\begin{eqnarray}\label{vsdvbs4}
\textrm{(\ref{vsdvbs3})}
\!\! &=& \!\!
\frac{-i}{\pi}
\lim_{\lambda \to 0}
\frac{1}{i\varepsilon-i\lambda}\,
\frac{\pi^2}{2u \sinh(\pi u)}+{\cal O}(\varepsilon^2)
\nonumber\\*[1.5mm]
\!\! &=& \!\!
\delta(u-v)\,\frac{\pi^2}{2u \sinh(\pi u)}+{\cal O}(\varepsilon^2).
\end{eqnarray}
From this result and the result of (\ref{rnast}) for the case of $u\not=v$, (\ref{wreh11}) is obtained. 


\end{document}